\documentclass{aastex631}

\usepackage{graphicx}	
\usepackage{amsmath,bm,amssymb}	
\usepackage{multirow, threeparttable, booktabs, makecell}

\usepackage{hyperref}

\begin{document}

\title[Mock generation]{Fast generation of mock galaxy catalogues with COLA}

\correspondingauthor{Xiaolin Luo; Le Zhang; Xiao-Dong Li}
\email{luoxl23@sjtu.edu.cn; zhangle7@mail.sysu.edu.cn; lixiaod25@mail.sysu.edu.cn}

\author{Jiacheng Ding}
\affiliation{School of Physics and Astronomy, Sun Yat-Sen University, Zhuhai 519082, P. R. China}
\affiliation{Peng Cheng Laboratory, No. 2, Xingke 1st Street, Shenzhen 518000, P. R. China}
\affiliation{CSST Science Center for the Guangdong–Hong Kong–Macau Greater Bay Area, SYSU, Zhuhai 519082, P. R. China}

\author{Shaohong Li}
\affiliation{School of Physics and Astronomy, Sun Yat-Sen University, Zhuhai 519082, P. R. China}
\affiliation{CSST Science Center for the Guangdong–Hong Kong–Macau Greater Bay Area, SYSU, Zhuhai 519082, P. R. China}

\author{Yi Zheng}
\affiliation{School of Physics and Astronomy, Sun Yat-Sen University, Zhuhai 519082, P. R. China}
\affiliation{CSST Science Center for the Guangdong–Hong Kong–Macau Greater Bay Area, SYSU, Zhuhai 519082, P. R. China}

\author{Xiaolin Luo}
\affiliation{School of Physics and Astronomy, Shanghai Jiao Tong University, Shanghai 200240, P. R. China}

\author{Le Zhang}
\affiliation{School of Physics and Astronomy, Sun Yat-Sen University, Zhuhai 519082, P. R. China}
\affiliation{Peng Cheng Laboratory, No. 2, Xingke 1st Street, Shenzhen 518000, P. R. China}
\affiliation{CSST Science Center for the Guangdong–Hong Kong–Macau Greater Bay Area, SYSU, Zhuhai 519082, P. R. China}

\author{Xiao-Dong Li}
\affiliation{School of Physics and Astronomy, Sun Yat-Sen University, Zhuhai 519082, P. R. China}
\affiliation{Peng Cheng Laboratory, No. 2, Xingke 1st Street, Shenzhen 518000, P. R. China}
\affiliation{CSST Science Center for the Guangdong–Hong Kong–Macau Greater Bay Area, SYSU, Zhuhai 519082, P. R. China}

\begin{abstract}
We investigate the feasibility of using COmoving Lagrangian Acceleration ($\mathtt{COLA}$) technique to efficiently generate galaxy mock catalogues that can accurately reproduce the statistical properties of observed galaxies. Our proposed scheme combines the subhalo abundance matching (SHAM) procedure with $\mathtt{COLA}$ simulations, utilizing only three free parameters: the scatter magnitude ($\sigma_{\rm scat}$) in SHAM, the initial redshift ($z_{\rm init}$) of the $\mathtt{COLA}$ simulation, and the time stride ($da$) used by $\mathtt{COLA}$. In this proof-of-concept study, we focus on a subset of BOSS $\mathtt{CMASS~NGC}$ galaxies within the redshift range $z\in [0.45, 0.55]$. We perform $\mathtt{GADGET}$ simulation and low-resolution $\mathtt{COLA}$ simulations with various combinations of $(z_{\rm init}, da)$, each using $1024^{3}$ particles in an $800~h^{-1}{\rm Mpc}$ box. By minimizing the difference between $\mathtt{COLA~mock}$ and $\mathtt{CMASS~NGC}$ galaxies for the monopole of the two-point correlation function (2PCF), we obtain the optimal $\sigma_{\rm scat}$. We have found that by setting $z_{\rm init}=29$ and $da=1/30$, we achieve a good agreement between $\mathtt{COLA~mock}$ and $\mathtt{CMASS~NGC}$ galaxies within the range of 4 to $20~h^{-1}{\rm Mpc}$, with a computational cost two orders of magnitude lower than that of the  N-body code. Moreover, a detailed verification is performed by comparing various statistical properties, such as anisotropic 2PCF, three-point clustering, and power spectrum multipoles, which shows similar performance between $\mathtt{GADGET~mock}$ and $\mathtt{COLA~mock}$ catalogues with the $\mathtt{CMASS~NGC}$ galaxies. Furthermore, we assess the robustness of the $\mathtt{COLA~mock}$ catalogues across different cosmological models, demonstrating consistent results in the resulting 2PCFs. Our findings suggest that $\mathtt{COLA}$ simulations are a promising tool for efficiently generating mock catalogues for emulators and machine learning analyses in exploring the large-scale structure of the Universe.
\end{abstract}

\keywords{methods: data analysis, numerical --- cosmology: large-scale structure of Universe, theory}

\section{Introduction}
\label{sec:introduction}
The observable universe provides a wealth of information concerning the evolution of cosmology, and the statistics of its large-scale structure (LSS) plays a crucial role in constraining cosmological models. The study of LSS relies on large-scale sky surveys, which can be classified into three main types, i.e. photometric and spectroscopic galaxy surveys, and line-intensity mapping (LIM) surveys. The advancement of precision cosmology has been greatly aided by the availability of precise observational data, coupled with the progress in cosmological simulations and statistical methods. 

In recent years, the advancement in Large-Scale Structure (LSS) analysis has required the use of considerably larger sets of simulations. Currently, the prevailing statistical estimators utilized to investigate LSS include the two-point correlation function (2PCF)~\citep{landy1993bias}, the three-point correlation function (3PCF)~\citep{szapudi1998new}, and Fourier space multipoles~\citep{feldman1993power,hand2017optimal}. In order to accurately model the relationship between these statistics and cosmological parameters, it becomes imperative to rely on emulators constructed using extensive sets of simulations~\citep{euclid2019euclid,euclid2021euclid,mcclintock2019aemulus,zhai2019aemulus}. Moreover, in recent years machine learning algorithms have increasingly found diverse applications in the analysis of galaxy survey data~\citep{Ravanbakhsh:2017bbi, Schmelzle:2017vwd, Ntampaka:2019ole, Li2020...ML...2020SCPMA..63k0412P,  AIBAORecon...2020arXiv200210218M, 2021JCAP...09..039L, Wu:2021jsy, anagnostidis2022cosmology, makinen2022cosmic, Wu:2023wmj}. Therefore, acquiring mock catalogues of galaxies in different cosmologies becomes indispensable for effectively training machine learning models and constraining cosmological parameters. Hence, the generation of ample mock catalogs in different cosmologies is extremely important, irrespective of whether one chooses to employ manually designed statistics as the data summary or adopts machine-learning methods.

The advancement of observations also necessitates extensive simulations. In the next decade, the stage IV surveys such like DESI~\citep{2016arXiv161100036D}, EUCLID~\citep{2011arXiv1110.3193L}, Roman~\citep{2015arXiv150303757S}, LSST~\citep{2009arXiv0912.0201L}, CSST~\citep{2011SSPMA..41.1441Z, zhanhu2021, miao2023cosmological}, Subaru~\citep{2018PASJ...70S...4A} and SKA~\citep{2019arXiv191212699B,2020PASA...37....2W} are going to map large areas of the sky with unprecedented accuracy and efficiency, resulting in vast amounts of data. Consequently, the development of rapid mock generation techniques becomes crucial in supporting cosmological research that utilizes data from these new-generation sky surveys.

In order to generate mock catalogues that can be compared to observations, it is crucial to employ methods that establish a connection between the distribution of galaxies and the underlying dark matter field. Within the context of LSS analysis, the two traditional employed approaches are the Halo Occupation Distribution (HOD) and the Subhalo Abundance Matching (SHAM) methods. In recent years, several new methods have been proposed. For example,  \cite{behroozi2019universemachine, behroozi2020universe} proposed a method to flexibly and self-consistently determine individual galaxies’ star formation rates from their host haloes’ potential well depths, assembly histories, and redshifts. In 2022, \cite{wechsler2022addgals}  proposed the ADDGALS technique, which places galaxies within cosmological simulation lightcone outputs, yielding realistic mock galaxies. These mocks are generated using semi-analytic models from numerical simulations~\citep{yung2022semi,yung2023semi,bose2022constructing}. These methods play a vital role in bridging the gap between observed galaxy distributions and the underlying dark matter structures. By establishing a connection between galaxies and their corresponding dark matter halos, these techniques not only facilitate follow-up cosmological analysis, but also foster a comprehensive understanding of the LSS and its relationship with galaxy properties.

The concept of HOD originated in the early 2000s with pioneering studies such as \cite{jing1998spatial,peacock2000halo,seljak2000analytic}. Subsequently, it was further refined and explored in \cite{kravtsov2004dark}, which specifies the probability distribution for the number of galaxies that meet certain criteria (such as a luminosity or stellar mass threshold) within a halo, conditioned on its mass, denoted as $P(N|M)$. Building upon this, an extended version of the HOD model was proposed by \cite{yang2003constraining}, which establishes a connection between the full distribution of galaxy luminosity and the distribution of DM halos based on the conditional luminosity function. Notably, \cite{yang2005halo} utilized this model to calibrate galaxy group finders in magnitude-limited redshift surveys. While the HOD model typically requires 3-5 parameters for a given galaxy sample \cite{zheng2005theoretical,reddick2013connection}, the functional form of the HOD can become complex for large galaxy samples analysis, and determining the optimal parameter values to generate mock samples that match with observed statistics can be a time-consuming task.

The SHAM method is grounded on a straightforward assumption that more luminous (or massive) galaxies are hosted by more massive halos. Implementing SHAM entails establishing a mapping between galaxy stellar mass and various halo properties, resulting in the same stellar mass functions (SMFs) but distinct clustering signals. Specifically, SHAM relies on matching the stellar mass of a galaxy ($M_*$) with the mass of the (sub)halo ($M_h$). However, a challenge arises from the fact that subhalo masses are subject to intense tidal stripping upon entering larger halos, rendering them unreliable indicators of subhalo size. To address this issue, \cite{kravtsov2004dark} proposed employing the maximum circular velocity of subhalos, denoted as $V_{\rm max}$, as a more robust property for matching galaxies. Furthermore, \cite{shu2012evolution} suggested that the relationship between galaxies and DM halos is not strictly one-to-one due to inherent physical scatter. Unlike the HOD method, the SHAM approach is largely non-parametric, aside from the scatter factor (as described in Sect.~\ref{subsec:SHAM}). This basic and intuitive assumption allows for faster and simpler generation of mock galaxy catalogs.

Currently, robust N-body simulation techniques utilize methods such as Particle-Particle Particle-Mesh (${\rm P^3M}$) or Tree-PM to precisely calculate the gravitational forces acted on each particle, resulting in a typical force resolution of approximately 0.001 $h^{-1}\rm Mpc$. For instance, the $\mathtt{GADGET}$ N-body/SPH simulation code, which is freely available~\citep{springel2001gadget, springel2005cosmological, springel2021simulating}, is based on the Tree-PM algorithm. However, these techniques have the drawback of requiring numerous timesteps to faithfully simulate structure formation across both large and small scales, thus demanding substantial computational resources.

In contrast, fast mock generation methods offer the capability to simulate the distribution of galaxies in the Universe with significantly reduced computational costs. Currently proposed methods include $\mathtt{PINOCCHIO}$~\citep{Monaco_2002}, $\mathtt{PTHalos}$~\citep{Scoccimarro_2002}, $\mathtt{QPM}$~\citep{10.1093/mnras/stt2071}, $\mathtt{PATCHY}$~\citep{Kitaura_2013}, $\mathtt{HALOGEN}$~\citep{Avila_2015}, $\mathtt{COLA}$~\citep{Tassev_2013}, and so on. These methods employ Lagrangian Perturbation Theory (LPT) or the Particle-Mesh (PM) method to simulate the evolution of dark matter particles, followed by halo-finding algorithms or halo bias models to generate mock galaxy catalogues that exhibit statistical properties consistent with observations. These methods provide an efficient alternative for generating mock catalogs while maintaining fidelity to the observed statistics.  Among them, $\mathtt{COLA}$ is a quasi-N-body method that places particles in a co-moving frame following LPT to simulate the large-scale dynamical evolution, and also uses a full-blown N-body code with the PM algorithm to compute the small-scale dynamics. This distinctive design allows $\mathtt{COLA}$ to generate mock catalogues efficiently while maintaining a commendable level of precision in the non-linear clustering regime. As a result, $\mathtt{COLA}$ emerges as a promising tool for fulfilling the requirements of mock generation tasks essential for emulator or machine learning analyses.

The $\mathtt{COLA}$ method has garnered significant attention in follow-up studies across various disciplines. For instance, \cite{2015arXiv150207751T} proposed a spatial extension of the N-body $\mathtt{COLA}$ method, enabling zoom-in simulations. Furthermore, \cite{Koda_2016} developed a technique using $\mathtt{COLA}$ to generate mock catalogues that incorporate low-mass halos, demonstrating its capability to resolve both massive and low-mass halos. \cite{koda2016fast} employed $\mathtt{COLA}$ and HOD methods to create 600 mock galaxy catalogues for the WiggleZ Dark Energy Survey. Additionally, \cite{2018MNRAS.473.3051I} introduced the $\mathtt{ICE-COLA}$ method, which efficiently generates weak lensing maps and halo catalogues in the lightcone, offering a rapid and accurate solution for generating mock catalogues to model galaxy clustering observables. Most recently, \cite{ferrero2021dark} utilized $\mathtt{ICE-COLA}$ to produce halo lightcone catalogs and applied HOD methods to generate mock galaxy lightcone catalogues for DES Y3 samples. These studies highlight the diverse applications and advancements facilitated by the $\mathtt{COLA}$ method and its variants.


In this study, we explore the application of SHAM method to the outputs of the $\mathtt{COLA}$ fast simulation technique in order to construct mock galaxy catalogs that accurately reproduce the observed clustering properties. We present our fast mock generation method in Sect.~\ref{sec:Fast mock}, which includes a brief explanation of the $\mathtt{COLA}$ algorithm, the halo-finder method, and the SHAM technique. Additionally, the observational data utilized in this study are detailed in Sect.~\ref{sec:SDSS-III/BOSS DR12 CMASS samples}. The determination of the optimal parameter used in the SHAM technique is presented in Sect.~\ref{sect:determine_sigma}. In Sect.~\ref{sec:Estimation of mocks}, we compare the clustering properties of mock catalogues generated from $\mathtt{GADGET}$ and $\mathtt{COLA}$ simulations with observed galaxies. We further present the $\mathtt{COLA}$ results in different cosmologies in Sect.~\ref{subsec:Mock set}. Finally, we summarize our results and conclude in Sect.~\ref{sec:conclusions}. 

In this paper, we employ three different "base" cosmology, listed below, to prevent any confusion.
\begin{itemize}
    \item[1)] The assumed cosmological parameters ($\Omega_{m}=0.2951$, $\Omega_{\Lambda}=0.7049$, $\Omega_{b}=0.0468$, $w=-1.0$, $\sigma_8=0.80$, $n_s=0.96$, $h=0.6881$) listed in Tab.~\ref{tab:TestCosmo setting} were used for simulating DM particles to compare the $\mathtt{COLA}$ technique with $\mathtt{GADGET}$. This is consistent with the cosmological data from the Planck results ~\citep{ade2016planck}.

    \item[2)] In Sect.~\ref{subsec:mock production}, we adopt a base cosmology consistent with the WMAP 5-year data~\citep{komatsu2009five}, with the following parameters: $\Omega_{m}=0.26$, $w=-1.0$, $h=0.71$, in order to maintain consistency with the analysis presented in ~\cite{li2016cosmological}.
    
    \item[3)] In Sect.~\ref{subsec:Mock set}, we assessed the impact of different cosmologies on our proposed scheme by using the Planck 2015 cosmology~\citep{ade2016planck} with the following parameters: $\Omega_{m} = 0.31$, $w = -1.0$, $\sigma_{8} = 0.82$, as the basis for various cosmological models, as detailed in Tab.~\ref{tab:MultiCosmo setting}.
\end{itemize}

\section{Fast mock generation}
\label{sec:Fast mock}

\subsection{Observational Data}
\label{sec:SDSS-III/BOSS DR12 CMASS samples}
The Baryon Oscillation Spectroscopic Survey (BOSS) is a part of the SDSS-III~\citep{bolton2012spectral, dawson2012baryon, eisenstein2011sdss}. BOSS aims to detect the characteristic scale imprinted by baryon acoustic oscillations (BAO) in the early universe by measuring the spatial distribution of luminous red galaxies (LRGs) and quasars~\citep{eisenstein2001spectroscopic}. The high-mass end of SMF is represented by Luminous Red Galaxies (LRGs), making them an ideal group to reproduce using SHAM.

BOSS provides redshift information for approximately $1.5$ million galaxies in a sky region of $\sim 10^{4}$ square degrees, divided into two samples: LOWZ and CMASS. The LOWZ sample comprises the brightest and reddest LRGs at $z\leq$0.4, while the CMASS sample targets galaxies at higher redshifts, many of which are also LRGs. The sky region covered by CMASS NGC galaxies is roughly $\rm RA\in[100^{\circ},270^{\circ}]$, $\rm DEC\in[-10^{\circ},70^{\circ}]$. 

For the purposes of our study, we focus solely on a subset of the BOSS DR12 CMASS NGC galaxies with a redshift range of $z\in [0.45, 0.55]$, simply labelled as $\mathtt{CMASS~NGC}$. This enables us to rapidly estimate the accuracy of the mock generation.

\subsection{COLA simulation}
\label{subsec:COLA simulation}
In N-body simulations the equation of motion is often solved using the standard leapfrog Kick-Drift-Kick algorithm. This algorithm discretizes the time evolution operator by applying a set of Kick and Drift operators, given by 
\begin{align}
    &\bm{x}_{i+1} = \bm{x}_i+\bm{v}_{i+1/2}\Delta t\,, \nonumber \\
    &\bm{v}_{i+1/2} = \bm{v}_{i-1/2}-\nabla \phi\Delta t\,,
\end{align}
Here, $\bm{x}_i$ ($i=0,1,2,\cdots$) denotes the position of a particle at time $t_i\equiv i \Delta t$, $\bm{v}_{i+1/2}$ is the velocity at $t_{i+1/2}\equiv (i+1/2)\Delta t$, and $\phi$ represents the gravitational potential.

The leapfrog process is a second-order accurate method for discretizing time, but its accuracy degrades for large time steps due to the truncation error from higher-order terms. To accurately integrate cosmological simulations, the time step is typically chosen to be proportional to the Hubble time $H^{-1}(t)$. However, at high redshifts where the Hubble time is small, the time step must be correspondingly small to avoid additional errors.

$\mathtt{COLA}$\footnote{https://bitbucket.org/tassev/colacode/src/hg/}, a hybrid approach that combines the second order LPT (2LPT) and N-body algorithms, is an effective solution for simulating dark matter particles. Perturbation theory has been successful in describing large scales, allowing for the linear growth rate to substitute for time integration in N-body simulations. $\mathtt{COLA}$ leverages this by using a comoving frame with observers following trajectories calculated in the perturbation theory. This tradeoff between accuracy at small scales and computational speed is achieved without sacrificing accuracy at large scales.

In the framework of $\mathtt{COLA}$, particles evolve in a frame that is comoving with "LPT observers". The process begins with the computation of the initial conditions using 2LPT. Next, particles are evolved along their 2LPT trajectories and a residual displacement is added relative to the 2LPT path. This displacement is then integrated numerically using an N-body solver. Mathematically,
\begin{align}
    &\bm{x} = \bm{x}_{\rm LPT} + \bm{x}_{\rm res}\,, \nonumber        \\
    &\bm{v} = \dot{\bm{x}}_{\rm LPT} + \bm{v}_{\rm res}\,,\nonumber   \\
    &\bm{F}(\bm{x}) = \ddot{\bm{x}}_{\rm LPT} + \bm{F}_{\rm res}(\bm{x})\,,
\end{align}
where $\bm{F}$ represents the force on the particle, and the dots represent time derivatives. In LPT, the Eulerian final comoving positions $\bm{x}$ are related to the initial positions $\bm{q}$ through a displacement field $\bm{\mathit{\Psi}}$:
\begin{align}
    \bm{x}(t)=\bm{q}+\bm{\mathit{\Psi}}(\bm{q},t)\,.
\end{align}
If applying 2LPT to the calculation, i.e., $\bm{x}_{\rm LPT}\rightarrow\bm{x}_{\rm 2LPT}$, then the residual displacement field $\bm{\mathit{\Psi}}_{\rm res}$ is given by,
\begin{align}
    \bm{\mathit{\Psi}}_{\rm res}(\bm{q},t) =  \bm{\mathit{\Psi}}(\bm{q}) - D_{1}(t)\bm{\mathit{\Psi}}_{1} (\bm{q}) - D_{2}(t)\bm{\mathit{\Psi}}_{2}(\bm{q})\,,
\end{align}
where $\bm{\mathit{\Psi}}_{1}$ and $\bm{\mathit{\Psi}}_{2}$ are the Zel'dovich and 2LPT displacement fields at the present day ($a=1$). 

Let us define the time operator $\mathcal{T}$ as 
\begin{align}
    \mathcal{T} \equiv \frac{a}{H_{0}}\partial_{\eta}=a^{3}\frac{H(a)}{H_{0}}\partial_{a}\,,
\end{align}
where $\eta$ is the conformal time, $a$ is the scale factor, and $H$ is the Hubble parameter with $H_{0}$ being its value today. Then the operation of Kick-Drift-Kick algorithm in $\mathtt{COLA}$ is given by
\begin{align}
    \label{eq:x_dick_cola}
    &\bm{x}_{i+1} = \bm{x}_i+\bm{v}_{i+1/2}\Delta t + \Delta D_{1}\bm{\mathit{\Psi}}_{1} + \Delta D_{2}\bm{\mathit{\Psi}}_{2}\,,   \\ 
    \label{eq:v_kick_cola}
    &\bm{v}_{i+1/2} = \bm{v}_{i-1/2}-\mathcal{T}^{2}[\bm{\mathit{\Psi}}_{{\rm res},i}]\Delta t\,,
\end{align}
where 
\begin{align}
    \mathcal{T}^{2}[\bm{\mathit{\Psi}}_{\rm res}]+\mathcal{T}^{2}[D_{1}]\bm{\mathit{\Psi}}_{1}+\mathcal{T}^{2}[D_{2}]\bm{\mathit{\Psi}}_{2}+\nabla\phi=0\,.
\end{align}
Here $\Delta D_{ n} = D_{{ n},i+1}-D_{{\rm n},i}$ for $n=1, 2, \cdots$, which denotes the change in the first and second-order growth factors over the timestep. $\mathtt{COLA}$ uses Eq.~\ref{eq:x_dick_cola} \&~Eq.~\ref{eq:v_kick_cola} to update the N-body particle positions and velocities, as well as to interpolate the quantities between timesteps for snapshots at redshifts of interest. Unlike the standard N-body methods, $\mathtt{COLA}$ relies solely on the PM method and 2LPT, leading to an imprecise force resolution and correspondingly a coarse resolution on small scales. For this reason, the force resolution of $\mathtt{COLA}$ simulations needs to be carefully considered when using the halo-finder.

To identify halos, we use the ROCKSTAR\footnote{https://bitbucket.org/gfcstanford/rockstar/src/main/} (Robust Overdensity Calculation using K-Space Topologically Adaptive Refinement) halo finder~\citep{behroozi2012rockstar}, which employs adaptive hierarchical refinement of Friend-of-Friend groups in six phase-space dimensions and one time dimension. The resulting halo samples from ROCKSTAR include both host halos and subhalos. To mitigate the small-scale uncertainty of $\mathtt{COLA}$, we adjust the force resolution in the ROCKSTAR halo-finder to approach the grid size in our $\mathtt{COLA}$ simulation. The modern SHAM method not only requires high-resolution simulations with resolved substructure but also necessitates accurate merger trees to track the path of halos. The evolution history from the merger tree is helpful for finding representative halos throughout the history of the Universe. Once ROCKSTAR creates the particle-based merger trees, we use the ``consistent trees'' algorithm~\citep{behroozi2012gravitationally}, to trace the evolution of halos with redshift. 

To assess the feasibility of using $\mathtt{COLA}$, we ran simulations with $1024^{3}$ particles in a box measuring 800 $h^{-1}{\rm Mpc}$, allowing for comparison with $\mathtt{GADGET~mock}$ under identical conditions. The initial matter power spectrum and transfer function were generated using CAMB~\citep{lewis2000efficient}. Subsequently, in Sect.~\ref{subsec:Mock set} we employed the best simulation settings for production of the mock catalogues. A detailed outline of the simulation settings can be found in Tabs.~\ref{tab:TestCosmo setting} \& \ref{tab:MultiCosmo setting}.

\begin{table}
\renewcommand\arraystretch{1.4}
\centering
    \begin{tabular}{c|cc|rrcrr|cc}
    \hline
        SimType & $\begin{array}{c}{\rm Box~size} \\ (h^{-3}{\rm Mpc}^3)\end{array}$ & $N_{\rm par}$ & $z_{\rm init}$ & $a_{\rm init}$ & $a_{\rm final}$ & $da$ & $N_{\rm step}$ & CPUs & $\begin{array}{c}{ \text{Wall-clock}} \\ \text{time} \;\; (\rm hr) \end{array}$    \\ \hline
        
        $\mathtt{GADGET}$-2 & $800^3$ & $1024^3$ & $49$  & 1/50  & 1 & $\mathbf{-}$ & 3676 & 252 & 149.7 \\ \hline 

        $\mathtt{COLA}$     & $800^3$ & $1024^3$ & $29$  & 1/30  & 1 & 1/30  & 30  & 256 & 0.39 \\ \hline
        $\mathtt{COLA}$     & $800^3$ & $1024^3$ & $49$  & 1/50  & 1 & 1/50  & 50  & 256 & 0.75 \\ \hline
        $\mathtt{COLA}$     & $800^3$ & $1024^3$ & $59$  & 1/60  & 1 & 1/60  & 60  & 256 & 0.76 \\ \hline
        $\mathtt{COLA}$     & $800^3$ & $1024^3$ & $119$ & 1/120 & 1 & 1/120 & 120 & 256 & 1.51 \\ \hline
        $\mathtt{COLA}$     & $800^3$ & $1024^3$ & $29$  & 1/30  & 1 & 1/120 & 117 & 256 & 1.37 \\ \hline
        $\mathtt{COLA}$     & $800^3$ & $1024^3$ & $59$  & 1/60  & 1 & 1/120 & 119 & 256 & 1.52 \\ \hline
        $\mathtt{COLA}$     & $800^3$ & $1024^3$ & $119$ & 1/120 & 1 & 1/120 & 120 & 256 & 1.51 \\ \hline

        $\mathtt{COLA}$     & $800^3$ & $1024^3$ & $29$  & 1/30  & 1 & 1/30  & 30  & 256 & 0.39 \\ \hline
        $\mathtt{COLA}$     & $800^3$ & $1024^3$ & $29$  & 1/30  & 1 & 1/60  & 59  & 256 & 0.78 \\ \hline
        $\mathtt{COLA}$     & $800^3$ & $1024^3$ & $29$  & 1/30  & 1 & 1/120 & 117 & 256 & 1.37 \\ 
         
    \hline 
    \end{tabular}
\caption{Simulation settings and the computational costs for the $\mathtt{GADGET}$-2 and $\mathtt{COLA}$ simulations. In both simulations, the same cosmological parameters were applied, specifically: $\Omega_{m} = 0.2951$, $\Omega_{\Lambda} = 0.7049$, $\Omega_b=0.0468$, $w=-1.0$, $\sigma_8=0.80$, $n_s=0.96$, and $h=0.6881$. The number of DM particles and time steps in simulation are denoted by $N_{\rm par}$ and $N_{\rm step}$, respectively. The redshift range in the simulations is determined by the initial redshift $z_{\rm init}$ and the final redshift $z_{\rm final}$. We will evaluate the performance of $\mathtt{COLA}$ by using various values listed for the initial redshift ($z_{\rm init}$), the time stride ($da$), and $N_{\rm step}$ of $\mathtt{COLA}$. When comparing the total wall-clock time of $\mathtt{GADGET}$ to that of $\mathtt{COLA}$, with a similar number of CPUs, it becomes evident that the computational cost of $\mathtt{GADGET}$ simulations is approximately $10^2$ times greater than that of $\mathtt{COLA}$ simulations.}
\label{tab:TestCosmo setting}
\end{table}

\begin{figure}
    \centering
    \includegraphics[width=0.5\columnwidth]{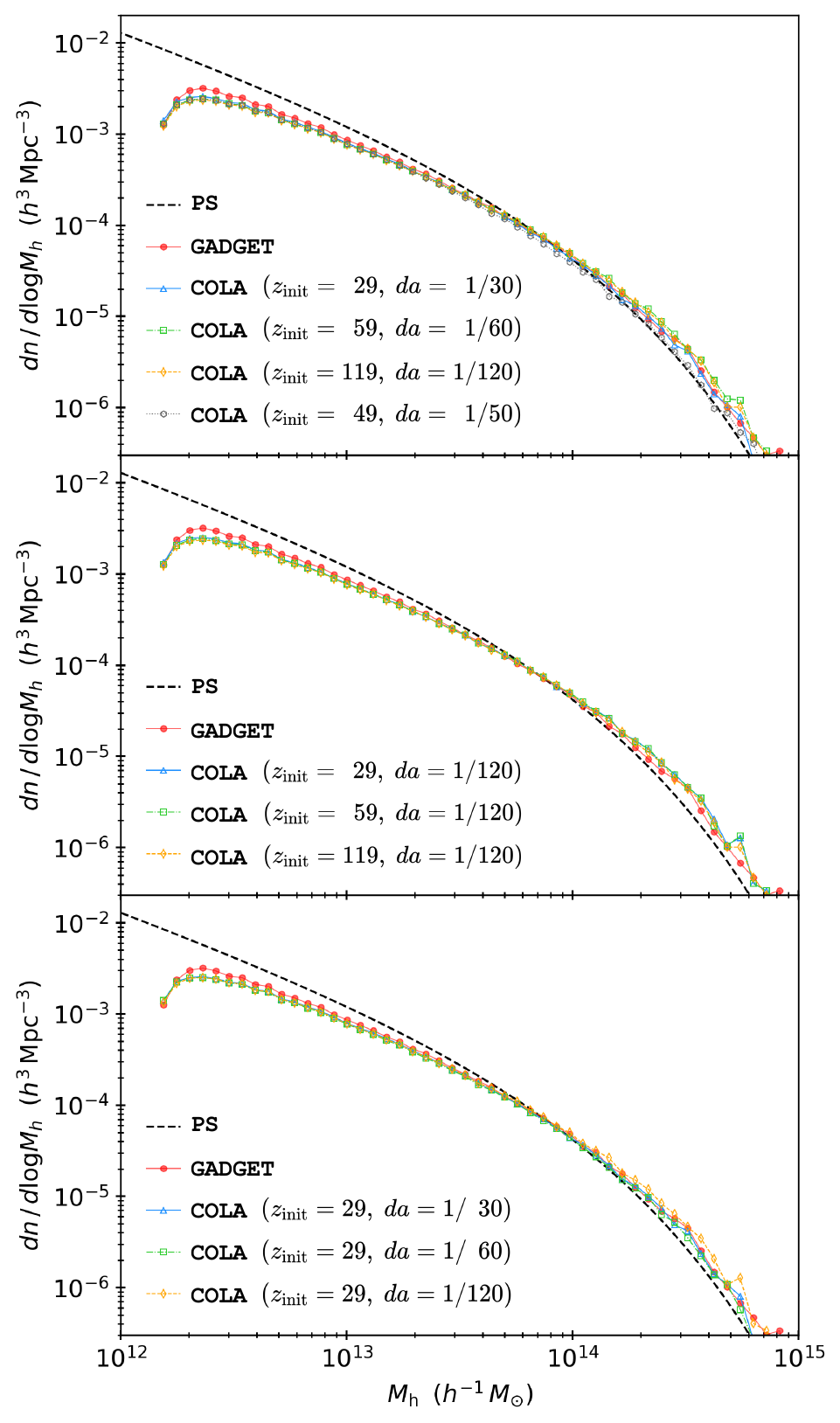}
    \caption{Comparison of halo mass functions (HMFs) of the simulated halos provided by $\mathtt{GADGET}$ simulation and $\mathtt{COLA}$ simulations with different settings of ($z_{\rm init}$, $da$), respectively. The theoretical Press-Schechter mass function, labeled as 'PS', is also shown as a black dashed line in each panel. As observed, the HMF derived from $\mathtt{COLA}$ agree well with the theoretical prediction and that derived from $\mathtt{GADGET}$. Furthermore, the results from different $\mathtt{COLA}$ parameter settings converge. In the low-mass region ($M_h \lesssim 2 \times 10^12$  $h^{-1}M_{\odot}$), the simulated HMFs in both  and $\mathtt{COLA}$ begin to noticeably deviate from the theoretical predication. This deviation is due to the limitations of mass resolution in N-body simulations.}
    \label{fig:TestSim_HMF}
\end{figure}

Furthermore, we perform a convergence test in the halo mass function (HMF) according to the simulated halos from  and $\mathtt{COLA}$ simulations respectively. The results are shown in Fig.~\ref{fig:TestSim_HMF}. As seen, both HMF of $\mathtt{GADGET}$ and HMFs of $\mathtt{COLA}$ simulations with different combinations of ($z_{\rm init}$, $da$) lead to comparable HMFs with the theoretical prediction~\citep{press1974formation,bond1991excursion}. By varying $z_{\rm init}$ and $da$, the HMFs simulated with $\mathtt{COLA}$ are highly consistent with the HMF of $\mathtt{GADGET}$.

\subsection{Stellar mass function}
\label{subsec:SMF}
The simulated halo samples correspond to the total galaxy samples covering all stellar mass intervals. However, due to the selection process in measurements, the stellar mass function (SMF) of a single observational survey is incomplete. By combining several galaxy surveys, the complete SMF can be obtained and be described by a fitted function. Typically, the observed SMF of quiescent galaxies is fitted with a double Schechter function~\citep{weigel2016stellar}, while star-forming galaxies are often described using a single Schechter function (e.g.,~\cite{li2009distribution, peng2012mass, muzzin2013evolution}). The SMF is known to evolve with redshift, but \cite{mitchell2016evolution} indicates that the relationship between stellar mass and halo mass weakly evolves over the redshift range of $0<z<4$. In this work, we use the redshift-invariant stellar mass function (SMF) by fitting it into a segmented function within a specific stellar mass interval. Each segment of the SMF satisfies the single Press-Schechter formula, given by 
\begin{align}
    \label{eq: Press-Schechter SMF}
    \phi(\log M_{*}) \;  d \log M_{*} =&  \ln(10)\exp\left(-10^{\log M_{*} - \log M_c}\right)  \\ \nonumber
    & \times \phi_c\left(10^{\log M_{*}-\log M_c}\right)^{\alpha+1}~d\log M_{*}.
\end{align}
The fitting parameters ($\phi_c$, $\alpha$, $\log M_c$) were proposed by~\cite{rodriguez2016clustering} and are shown in Tab.~\ref{tab: params Press-Schechter SMF}.
\begin{table}
    \renewcommand\arraystretch{1.2}
    \centering
    \begin{tabular}{cccc}
\hline\hline
$\begin{array}{c}\text { Mass range } \\
{(M_{\odot})}\end{array}$ & $\begin{array}{c}\phi_c \\
{(\mathrm{Mpc}^3 \log M_{\odot}^{-1})}\end{array}$ & $\alpha$ & $\begin{array}{c}\log M_c {(M_{\odot})}\end{array}$ \\ \hline
$\log M_* \leq 11.00$ & $4.002 \times 10^{-3}$ & $-0.938$ & 10.76 \\
\hline
$\log M_*>11.00$ & $2.663 \times 10^{-3}$ & $-2.447$ & 11.42 \\
\hline\hline
\end{tabular}
\caption{Parameters of the Press–Schechter SMF. Here $\alpha$ and $\phi_c$ denote the slope and the normalization, and $M_c$ is the characteristic mass. }
\label{tab: params Press-Schechter SMF}
\end{table}

\begin{figure}
    \centering
    \includegraphics[width=0.6\columnwidth]{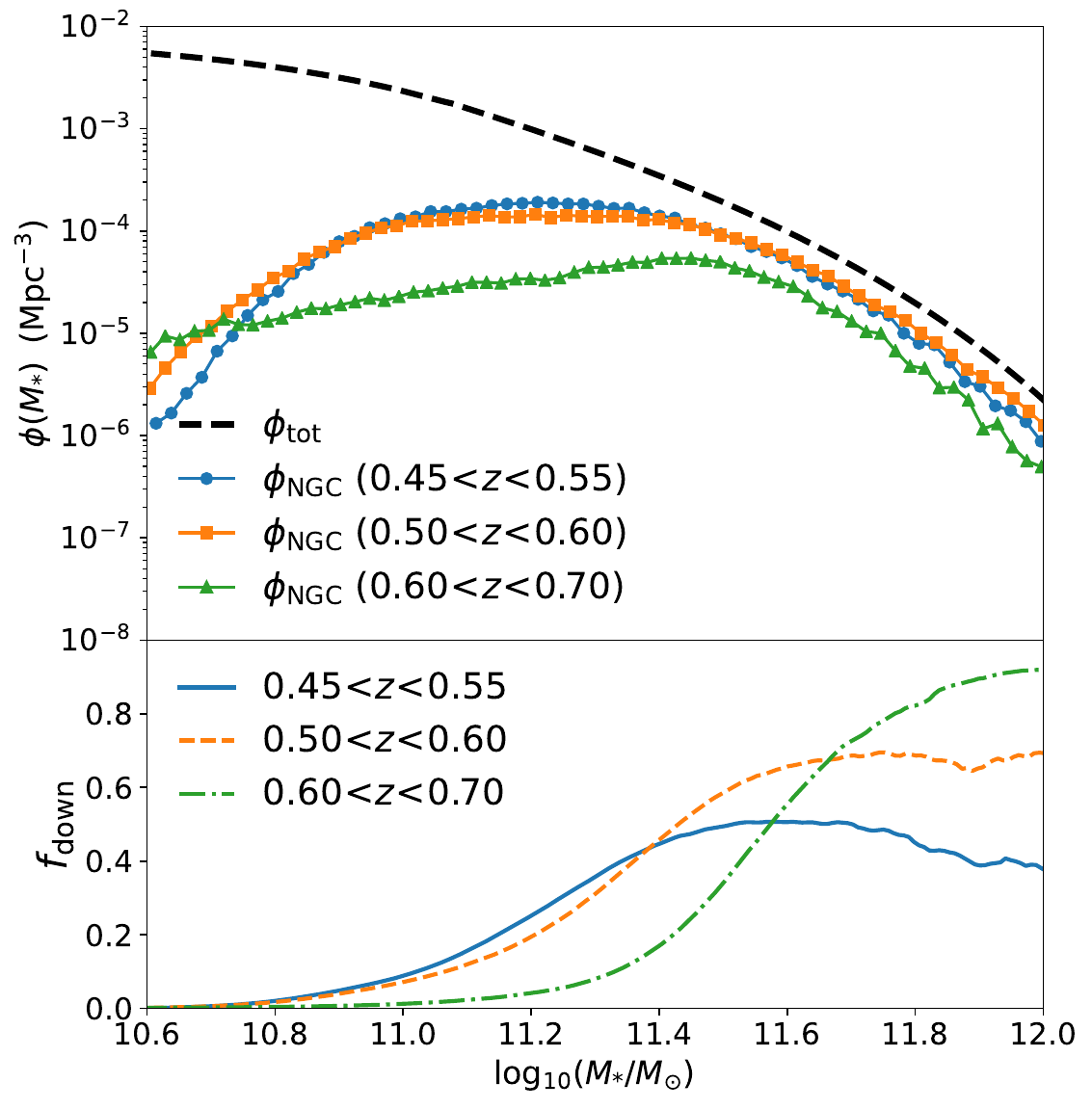}
    \caption{Top panel: the redshift-invariant complete SMF $\phi_{\rm tot}$ (dashed), as described in Sect.~\ref{subsec:SMF}, along with the observed SMFs of the $\mathtt{CMASS~NGC}$ catalogue in three redshift bins (marked) denoted by $\phi_{\rm NGC}$. Bottom panel: downsampling ratio $f_{\rm down}$ for three redshift bins. This ratio is defined as the number density of observed galaxies in a given bin ($\phi_{\rm NGC}$) divided by the number density of the full galaxy population ($\phi_{\rm tot}$). Since observations are inherently incomplete, the observed catalogue in each stellar mass bin represents a subset of the full galaxy population in the same sky region. Note that our analysis in this study specifically focuses on catalogues within the redshift range of $0.45<z<0.55$ (solid blue). }
    \label{fig:SMF-downsample}
\end{figure}

To match the stellar mass catalogues and the BOSS DR12 catalogues, we employ three key quantities, namely PLATE, MJD, and FIBERID, to uniquely identify each galaxy in the BOSS DR12 catalogues. For the $\mathtt{CMASS~NGC}$ catalogue, we combine the raw released catalogue with the Portsmouth SED-fit DR12 stellar mass catalogue~\citep{maraston2009modelling} to obtain the galaxy catalogue enriched with stellar mass. We then obtain the SMF of the $\mathtt{CMASS~NGC}$ catalogue. By combining the complete SMF and the SMF of the $\mathtt{CMASS~NGC}$ catalogue, we can calculate the downsampling ratios $f_{\rm down}$ for each mass bin, as illustrated in Fig.~\ref{fig:SMF-downsample}. These ratios are used to generate mock catalogues of the $\mathtt{CMASS~NGC}$ catalogue.

\subsection{Subhalo abundance matching procedure}
\label{subsec:SHAM}
SHAM is a simple and powerful statistical approach for connecting galaxies to subhalos. In its simplest form, given some property of subhalos, such as halo mass or maximum circular velocity, the subhalo number density and the galaxy number density are matched in order to obtain the connection between subhalos and the galaxies that they host. Some SHAM related works focus on fitting the parameters in a function describing the stellar mass vs. halo mass (SMHM) relation in order to minimize the deviation between the model SMF and an observed SMF~\citep{rodriguez2017constraining, moster2018emerge, behroozi2019universemachine}. 
 
In this study, our focus does not lie in determining the SMHM relation, since the $\mathtt{COLA}$ simulation may result in a non-negligible uncertainty of halo mass. Specifically, our implementation of the SHAM process can be divided into the following steps:
\begin{itemize}
    \item[(1)] For the host halos, we choose the maximum circular velocity, denoted as $V_{\rm max}$, which represents the highest circular speed attained by test particles within these halos. This property is considered as the halo property for the subhalo abundance matching (SHAM) technique. However, there is an additional complication caused by the significant evolution of subhalos within host halos due to interactions in the dense environments of the larger hosts. Consequently, using $V_{\rm max}$ of a subhalo as a proxy for stellar masses may not yield accurate results. To address this issue, it has become common practice to assign stellar masses or luminosities to subhalos based on their $V_{\rm peak}$ value, as proposed by~\cite{hearin2013sham}, where $V_{\rm peak}$ represents the largest $V_{\rm max}$ achieved by a subhalo throughout its entire history.

    \item[(2)] The assumption underlying SHAM is that more luminous (or massive) galaxies are typically found within more massive halos. However, this relationship is not strictly one-to-one, as there exists a natural scatter between galaxies and dark matter halos. To account for this scatter, we employ a method proposed by~\cite{rodriguez2016clustering},  for assigning stellar mass to (sub)halos. This involves defining a scattered quantity, denoted as $V^{\rm scat}_L$, and the expression is as follows: 
\begin{equation}
V_L^{\rm scat }=V_{L}\left[1+\mathcal{N}\left(0, \sigma_{\rm scat }\right)\right]\,,
\end{equation}
with 
\begin{equation}    \label{eq:V_scatter}
V_L= \begin{cases}V_{\rm max } & {\rm  (host~halos) } \\ V_{\rm peak } & {\rm (subhalos) }\end{cases}\,,
\end{equation}
where $\mathcal{N}$ is a random number following a Gaussian distribution with zero mean and a standard deviation of $\sigma_{\rm scat} (V_L| M_{*})$. 

\item[(3)] We connect the scattered quantity $V_L^{\rm scat}$ of (sub)halos to the stellar mass ($M_*$) of the central galaxies by assuming a monotonic relationship between them. If the number density of (sub)halos with $V_L^{\rm scat}$ matches the number density of galaxies with a stellar mass exceeding $M_*$, 
    \begin{align}
    n(>V_L^{\rm scat}) = n(>M_*)\,,
    \end{align}
    then we assume that galaxies with a stellar mass of $M_*$ reside at the center of (sub)halos with $V_L^{\rm scat}$. Our study uses a lower limit of stellar mass $\log_{10}M_{*}^{\rm cut}=11.0$.
\end{itemize}

\subsection{Mock generation}
\label{subsec:mock production}
In this study, we mainly generate mock catalogues in a cosmology using the two different N-body simulation codes. To validate the robustness and effectiveness of the $\mathtt{COLA}$ simulation, we examine three different cases with varying $\mathtt{COLA}$ settings. In the first case, we maintain the default settings of $\mathtt{COLA}$, i.e., varying $z_{\rm init}$ and adjusting $da$ according to $da=1/(1+z_{\rm init})$. In the second case, we vary $z_{\rm init}$ while keeping $da$ constant at $da=1/120$. In the final case, we keep $z_{\rm init}$ at 29 while altering $da$. Further details are provided in Tab.~\ref{tab:TestCosmo setting}.

The mock catalogues are labeled by their respective N-body method, $\mathtt{GADGET~mock}$ and $\mathtt{COLA~mock}$. Both simulations use $1024^{3}$ DM particles to trace the evolution and are conducted in cubic volumes with periodic boundary conditions and a side length of 800 $h^{-1}{\rm Mpc}$ creating 16 snapshots in the redshift range $0.2<z<0.7$. The particle mass corresponds to $3.9\times 10^{10}\,h^{-1}M_{\odot}$. For the $\mathtt{COLA}$ N-body method, the force resolution is determined by the simulated box size and the mesh size. In the default setting of $\mathtt{COLA}$ the mesh size is chosen as $3^3$ times larger than the number of DM particles ($N_{\rm par}$), resulting in a force resolution of approximately $260~h^{-1}{\rm kpc}$ in a cubic volume with $800^3~h^{-3}{\rm Mpc^3}$ for $N_{\rm par}=1024^{3}$. We then use ROCKSTAR halo-finder and the Merger-Tree technique to identify halos from the DM particle snapshots of the $\mathtt{COLA}$ simulation. The input force resolution of ROCKSTAR halo-finder is set as the force resolution of the $\mathtt{COLA}$ simulation. The output redshifts of snapshots from the $\mathtt{COLA}$ simulation were chosen as follows: $z_{\rm snap}=$ [0.71, 0.66, 0.62, 0.57, 0.53, 0.50, 0.46, 0.42, 0.39, 0.36, 0.33, 0.30, 0.27, 0.25, 0.22, 0.20]. These redshift values were selected in order to match the redshift range of the SDSS BOSS catalogues. 

Applying the SHAM procedure, we generate mock catalogues with statistical properties that match those of observed galaxies in the $\mathtt{CMASS~NGC}$ dataset. In this proof-of-concept study, we limit our analysis to a narrow redshift range of $0.45<z<0.55$ to enable a fast comparison between the statistical properties of the $\mathtt{COLA~mock}$ and $\mathtt{GADGET~mock}$ catalogues with those of $\mathtt{CMASS~NGC}$.

The overall procedure for generating mock galaxies from the snapshot of halos is as follows.
\begin{figure}
    \centering
    \includegraphics[width=0.6\columnwidth]{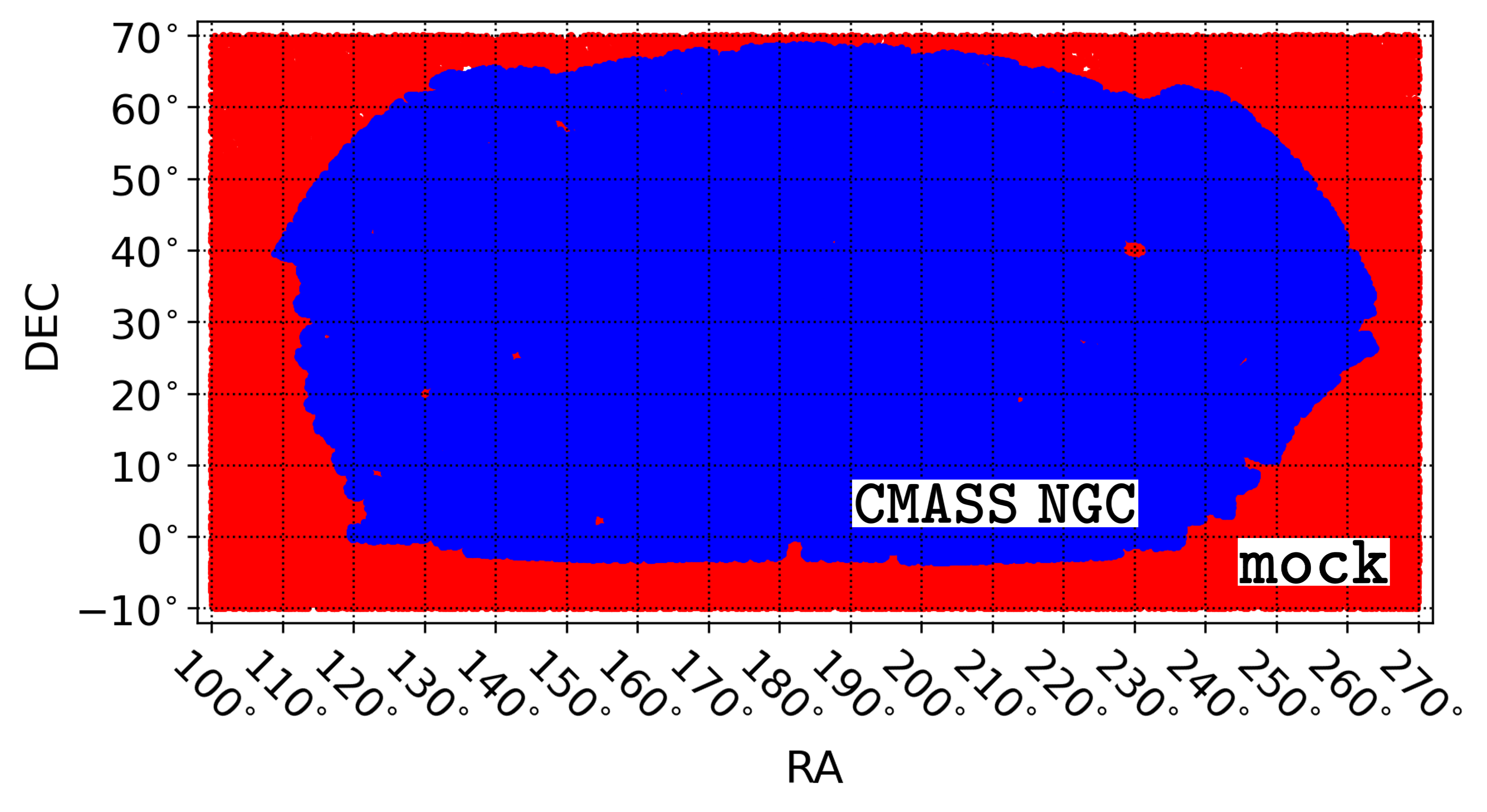}
\caption{Comparison of the sky coverage. The area shaded in blue represents the sky coverage of $\mathtt{CMASS~NGC}$. For simplicity, we do not apply the angular selection of $\mathtt{CMASS~NGC}$ to the mock catalogues ($\mathtt{GADGET}$ and $\mathtt{COLA~mocks}$) when estimating the two-point or three-point clustering. This means that the sky coverage of the mocks includes the entire region in red. However, when calculating the multipoles of the power spectrum (see Sect.\ref{sect.pk}), we do apply the angular selection of $\mathtt{CMASS~NGC}$ to the mock catalogues.}
    \label{fig:sky-region}
\end{figure}

\begin{itemize}
    \item[(1)] Using the redshift information available in the snapshots, we create periodic replicas of the snapshot of (sub)halos at $z=0.50$, and create a sample of objects in a redshift-shell which corresponds to the true redshift range $0.4 <z<0.6$. 
    
    \item[(2)]  To account for the effect of redshift space distortion (RSD) on the halo position within the redshift-shell, we use the following relation:
    \begin{align}
    z_{\rm obs} = z + (1+z)  \frac{v_{\parallel}}{c}\,,
    \end{align}
    where $z$ represents the real redshift without RSD, while $z_{\rm obs}$ represents the observed redshift after accounting for RSD. The projected velocity of the halo along line-of-sight (LoS) is denoted by $v_{\parallel}$, and $c$ represents the speed of light. We then update the redshift-shell satisfying the condition that $0.45<z_{\rm obs}<0.55$.
    
    \item[(3)] We select a block of sky region covering $\rm RA\in[100^{\circ},270^{\circ}]$, $\rm DEC\in[-10^{\circ},70^{\circ}]$ which corresponds to the sky coverage of $\mathtt{CMASS~NGC}$ galaxies (as shown in Fig.~\ref{fig:sky-region}). This selection is made without taking into account any veto masks or fibre collisions in the BOSS DR12 data.

    \item[(4)] Considering the complete SMF with the lower limit of stellar mass $\log_{10}M_{*}^{\rm cut}=11.0$, the number density of the chosen full galaxies can be obtained, then we apply the SHAM procedure to obtain the (sub)halo catalogue with the same number density. The assumption is that the mock galaxies located at the center of halos have a stellar mass distribution that follows the complete SMF ($\phi_{\rm tot}$) mentioned in Sect.~\ref{subsec:SMF}.
    
    \item[(5)] Downsampling is necessary due to the inherent incompleteness of observations. This is because the galaxies we observe are a subset of the full galaxy population in the same sky region. As a result, the observed SMF, such as $\phi_{\rm NGC}$ shown in Fig.~\ref{fig:SMF-downsample}, is lower than the complete SMF, $\phi_{\rm tot}$, described in Sect.~\ref{subsec:SMF}. The downsampling ratio $f_{\rm down}$ represents the ratio of $\phi_{\rm NGC}$ to $\phi_{\rm tot}$, as illustrated in Fig.~\ref{fig:SMF-downsample}.

    \item[(6)] By applying the above procedures, we can generate a set of halo catalogs with varying $\sigma_{\rm scat}$ values in the SHAM procedure. In order to quantify the differences between these halo catalogs and the observed $\mathtt{CMASS~NGC}$ galaxies, we use the two-point clustering statistic $\xi_0(s)$, as defined in Sect.~\ref{sect:determine_sigma}, and calculate the corresponding $\chi^2$ value, given by 
    \begin{align}
    \label{eq:chi2_for_HAM_best_scatter}
        \chi^{2} &= \sum_{i,j}\Delta \xi_{0}^{i} (\bm{C}^{-1})_{ij} \Delta \xi_{0}^{j}\,,\quad {\rm with}~~ \Delta \xi_{0}^{i} = \xi_{0}^{\rm mock}(s_i) - \xi_{0}^{\rm obs}(s_i)\,,     
    \end{align}
    where the $\Delta \xi^{i}_{0}$ represents the deviation of $\xi_{0}$ between the halo catalogues and the $\mathtt{CMASS~NGC}$ galaxies at the bin $s_i$, while the covariance matrix $\bm{C}$ is determined from a large set of $\mathtt{PATCHY~mocks}$ as defined in Eq.~\ref{eq:The covariance matrix from PATCHY mocks}. It is worth noting that in this paper we apply the unified fiducial cosmology of ($\Omega_{m}=0.26$, $w=-1.0$, $h=0.71$) to estimate the statistics of mock catalogs.
\end{itemize}

\subsection{MultiDark Patchy BOSS DR12 mock catalogues}
To accurately estimate the covariance matrix, we use 1,000 MultiDark Patchy BOSS DR12 mock catalogues~\citep{kitaura2016clustering}, labeled as $\mathtt{PATCHY~mock}$ in this paper. These mock catalogues are generated using approximate gravity solvers and  analytical-statistical biasing models, and they are calibrated to a BigMultiDark N-body simulation \citep{rodriguez2016clustering} that employs $3,840^3$ particles in a $(2.5~h^{-1}\rm{Gpc})^3$ volume. This simulation assumes a $\Lambda$CDM cosmology with $\Omega_{m} = 0.307$, $\Omega_{b} = 0.048$, $\sigma_8 = 0.82$, $n_s = 0.96$, and $h=0.67$. Applying the aforementioned technique in several redshift bins, the resulted mock catalogues can match the redshift evolution of the biased tracers in the BOSS observations, and finally the contiguous lightcone can be created by combining the resulting mocks in several redshift bins. The MultiDark Patchy BOSS DR12 mock catalogues that result from this method accurately reproduced the number density, selection function, and survey geometry of the BOSS DR12 data. The 2PCF of the observational data was reproduced down to a few Mpc scales, with most within $1\sigma$~\citep{kitaura2016clustering}. These mock surveys have been utilized in a series of studies for the statistical analysis of BOSS data~\citep{alam2017testing} and references therein). The extensive set of mock catalogues enabled us to perform a robust statistical error estimation.

In general, for a given observable in a particular bin $i$, say $\mathcal{O}(i)$, we can estimate the covariance matrix by computing the sample covariance of simulated mock catalogues through 
\begin{align}
\label{eq:The covariance matrix from PATCHY mocks}
\bm{C}_{ij} = \frac{1}{N_m-1} \sum^{N_m}_{m=1} \left[ \mathcal{O}_m(i)-\overline{\mathcal{O}}(i)  \right] \left[\mathcal{O}_m(j)-\overline{\mathcal{O}}(j)\right]\,,
\end{align}
where the sum is performed over $N_m$ mock catalogues, $\overline{\mathcal{O}}$ represents the mean value over those mocks, and the index $m$ denotes the $m$-th realization mock.

\section{Determination of the optimal value of scatter factor}\label{sect:determine_sigma}
In this section, we will present the results of determining the optimal scatter factor $\sigma_{\rm scat}$ of SHAM for different $N_{\rm step}$ in the COAL simulation. The optimal $\sigma_{\rm scat}$ is derived by minimizing the difference between the $\mathtt{COLA~mock}$ and $\mathtt{CMASS~NGC}$ through $\chi^2$, as defined in Eq.~\ref{eq:chi2_for_HAM_best_scatter}. In this study, we propose to calculate $\chi^2$ using the monopole of the 2PCF instead of other statistical quantities.

\subsection{Monopole of 2PCF}
The Landy-Szalay estimator~\citep{landy1993bias} is used to calculate the two-point correlation function (2PCF)~\citep{davis1983survey} for both our mock catalogues and observation data. The estimator is defined as follows:
\begin{align}
\label{eq:xi_s_mu}
\xi(s,\mu) = \frac{DD-2DR+RR}{RR}\,,
\end{align}
where $DD$, $DR$, and $RR$ represent the normalized pair counts for galaxy-galaxy, galaxy-random, and random-random samples, respectively. These counts are separated by a distance defined by $s\pm\Delta s$ and $\mu\pm\Delta\mu$. Here, $s$ is the distance between the pair, and $\mu= {\rm cos}(\theta)$, with $\theta$ being the angle between the line joining the pair of galaxies and the LoS direction to the target galaxy. The presented statistic measures the anisotropy of the clustering signal. The random catalogue is composed of unclustered points with a number density in redshift space that simulates the radial selection function of the observational data. In order to decrease the statistical variance of the estimator, we construct random catalogues that are ten times larger than the data catalogues. With using the $\mathtt{CUTE}$ code\footnote{https://github.com/damonge/CUTE}, we  calculate the correlation function $\xi(s,\mu)$ for both the $\mathtt{CMASS~NGC}$ galaxies and the mock catalogues within the redshift range of $0.45<z<0.55$. 

The monopole $\xi_0(s)$ in configuration space is obtained by integrating $\xi(s,\mu)$ over each $s$, as given by 
\begin{align}
    \label{eq:xi_s}
    \xi_0(s)=\int \xi(s, \mu) d\mu\,.
\end{align} 
Note that due to the large uncertainty of the correlation function at large clustering scales, we restrict the computation of the $\chi^2$ to small clustering scales of $s\in[4,20]~h^{-1}{\rm Mpc}$. A smaller value of $\chi^{2}$ indicates a better match with the observed galaxies.

\subsection{Fitting results}

\begin{figure*}
    \centering
    \includegraphics[width=0.9\columnwidth]{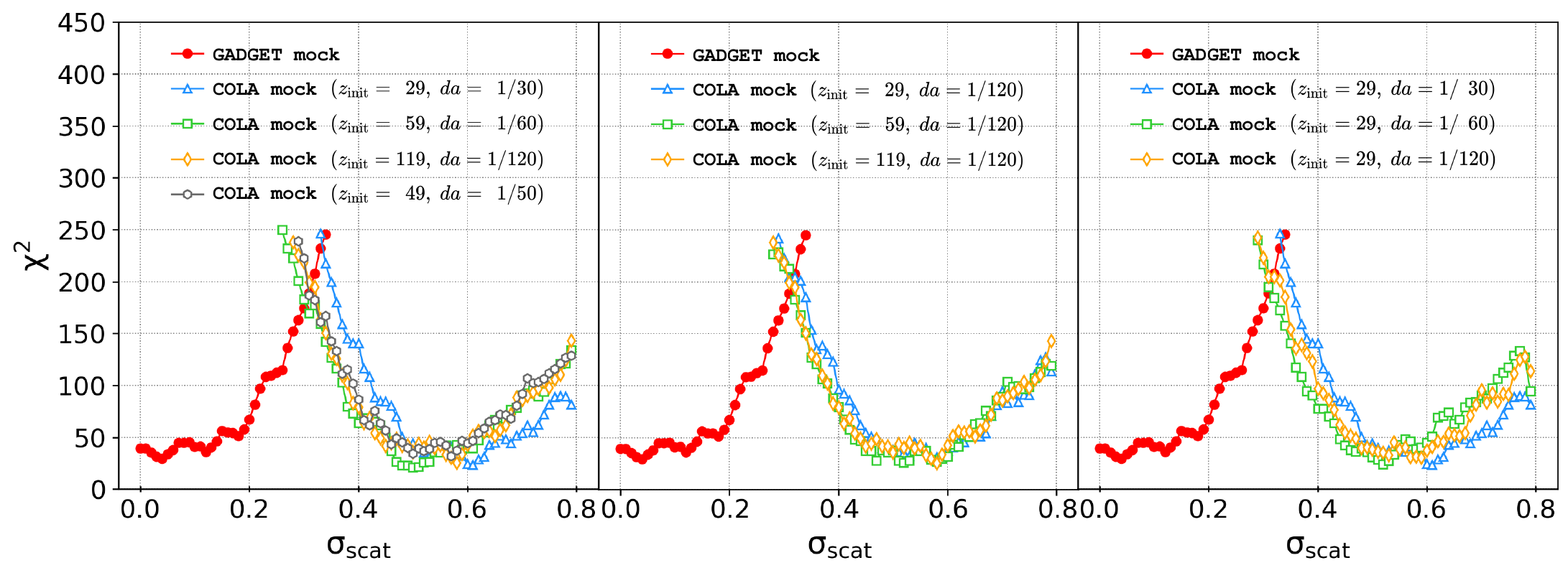}
    \caption{The $\chi^{2}$ estimation defined by Eq.~\ref{eq:chi2_for_HAM_best_scatter}, which is used to quantify the differences between mock catalogs and the observed CMASS NGC galaxies. We employ different $\mathtt{COLA}$ settings for comparison.}
    \label{fig:TestSim_chi2}
\end{figure*}

\begin{table} 
\centering
\renewcommand\arraystretch{1.2}
    \begin{tabular}{l|clcl|cc}
    \hline
        Catalogue & $z_{\rm init}$ & $a_{\rm init}$ & $a_{\rm final}$ & $da$ & $\sigma_{\rm scat}^{\rm best}$ & $\chi^{2}_{\rm min}$  \\ \hline
        $\mathtt{GADGET~mock}$  & 49  & 1/50  & 1 & $\mathbf{-}$ & 0.03  & 24.6  \\ \hline

        $\mathtt{COLA~mock}$    & 29  & 1/30  & 1 & 1/30  & 0.60  & 23.2  \\ \hline
        $\mathtt{COLA~mock}$    & 49  & 1/50  & 1 & 1/50  & 0.57  & 23.2  \\ \hline
        $\mathtt{COLA~mock}$    & 59  & 1/60  & 1 & 1/60  & 0.49  & 19.4  \\ \hline
        $\mathtt{COLA~mock}$    & 119 & 1/120 & 1 & 1/120 & 0.58  & 21.4  \\ \hline

        $\mathtt{COLA~mock}$    & 29  & 1/30  & 1 & 1/120 & 0.56  & 20.8  \\ \hline
        $\mathtt{COLA~mock}$    & 59  & 1/60  & 1 & 1/120 & 0.52  & 19.5  \\ \hline
        $\mathtt{COLA~mock}$    & 119 & 1/120 & 1 & 1/120 & 0.58  & 21.4  \\ \hline

        $\mathtt{COLA~mock}$    & 29  & 1/30  & 1 & 1/30  & 0.60  & 23.2  \\ \hline
        $\mathtt{COLA~mock}$    & 29  & 1/30  & 1 & 1/60  & 0.51  & 18.0  \\ \hline
        $\mathtt{COLA~mock}$    & 29  & 1/30  & 1 & 1/120 & 0.56  & 20.8  \\ 
    \hline
    \end{tabular}
    \caption{Best-fit values of $\sigma_{\rm scat}$ obtained by minimizing the $\chi^2$ values in Eq.~\ref{eq:chi2_for_HAM_best_scatter} for different simulations using the SHAM procedure. The optimal $\sigma_{\rm scat}$ is then used to determine the corresponding $\chi^{2}_{\rm min}$, which are shown in the rightmost column. The calculation of $\xi_0(s)$ is based on $s$ values ranging from 4 to $20~h^{-1}{\rm Mpc}$. In all cases, the DOF are 16, determined by the number of bins.}
    \label{tab:best_sigma_GADGET-COLA}
\end{table}
 
\begin{figure}
    \centering
    \includegraphics[width=\columnwidth]{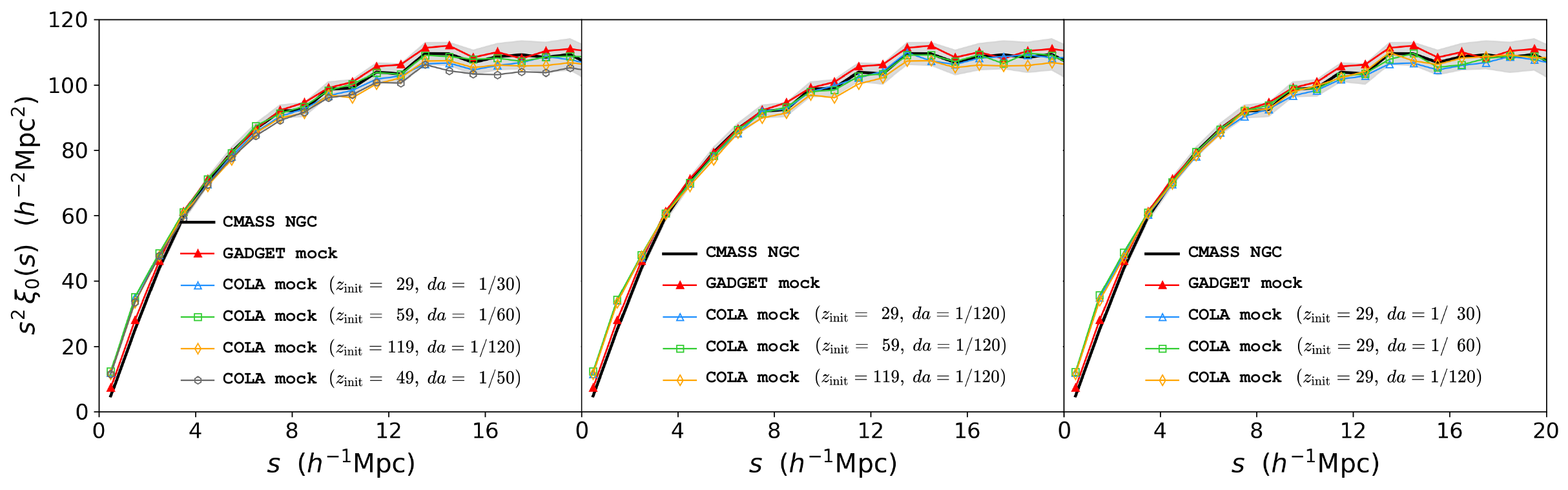}
    \caption{Comparison of the monopole of 2PCF for $\mathtt{GADGET~mock}$, $\mathtt{CMASS~NGC}$ and $\mathtt{COLA~mock}$ with different combinations of ($z_{\rm init}$, $da$). We observe a close match between $\xi_0(s)$ of $\mathtt{GADGET~mock}$ and $\mathtt{CMASS~NGC}$ within the range of $s\in[4,20]~h^{-1}{\rm Mpc}$. When $z_{\rm init}$ is set to 29, 49, and 59, the resulting $\xi_0(s)$ values of the $\mathtt{COLA~mocks}$ all fall within the $2\sigma$ region, indicating a good match with $\mathtt{CMASS~NGC}$.}
\label{fig:TestSim_SHAM_fit_2pcf}
\end{figure}

The resulting $\chi^2$ with varying $\sigma_{\rm scat}$ are presented in Fig.~\ref{fig:TestSim_chi2} and Tab.~\ref{tab:best_sigma_GADGET-COLA}. Based on the best scattering parameters ($\sigma_{\rm scat}^{\rm best}$), we obtained the mock catalogs for different simulations. Fig.~\ref{fig:TestSim_SHAM_fit_2pcf} displays the 2PCF curves at small scales for $\mathtt{GADGET~mock}$ and $\mathtt{COLA~mocks}$ with different combinations of ($z_{\rm init}$, $da$). It can be observed that the $\xi_0(s)$ of the $\mathtt{GADGET~mock}$ exhibits good consistency with that of $\mathtt{CMASS~NGC}$ within the selected scale range of $s\in[4, 20]~h^{-1}{\rm Mpc}$. 

In Fig.~\ref{fig:TestSim_SHAM_fit_2pcf}, we present the three cases. The left panel corresponds to $da = 1/(1+z_{\rm init})$, with $z_{\rm init}$ values of 29, 49, 59, and 119. In all cases, the $\xi_0(s)$ values fall within the $2\sigma$ region, indicating a good match with $\mathtt{CMASS~NGC}$. The middle panel represents the case where $z_{\rm init}$ varies while fixing $da = 1/120$, also showing good consistency with $\mathtt{CMASS~NGC}$. The right panel shows the case where $z_{\rm init}$ is fixed at 29, but $da$ is varied. In this case, the $\xi_0(s)$ values also fall within the $2\sigma$ region, indicating that different time strides $da$ in $\mathtt{COLA}$ lead to a similar evolution process of dark matter particles. 

Based on these results, we find that:
\begin{itemize}
\item For the $\mathtt{GADGET~mock}$, the minimum value of $\chi^2_{\rm min}=24.6$ is achieved with Degree-of-Freedom (DOF) $=16$ at $\sigma_{\rm scat}^{\rm best}=0.03$, which is consistent with the $\chi^2$ distribution at the $2\sigma$ level, since $\chi^2_{\rm min}< 27.3$, which is calculated as $16+2\sqrt{2\times16}$. These results indicate that the reference mock catalogue effectively reproduce the two-point clustering characteristics of the $\mathtt{CMASS~NGC}$ galaxies.
\item The $\mathtt{COLA}$ simulations for the case with $da=1/(1+z_{\rm init})$ and $z_{\rm init}$ set to 29, 59, 49 and 119, yield $\chi^2_{\rm min}$ values that consists with the $\chi^2$ distribution at the $2\sigma$ level ($\chi^2_{\rm min}<27.3$ when DOF=16). Specifically, the $\chi^2_{\rm min}$ values are 23.2, 23.2, 19.4 and 21.4 for the respective simulations, as listed in Tab.~\ref{tab:best_sigma_GADGET-COLA}. In the case where $z_{\rm init}=49$, the $\chi^2_{\rm min}$ of $\mathtt{COLA}$ with $N_{\rm step}$ is comparable to that of $\mathtt{GADGET~mock}$ with the same $z_{\rm init}$ but a significantly larger number of steps, $N_{\rm step}=3676$. This suggests that the $\mathtt{COLA}$ scheme can perform similarly to $\mathtt{GADGET}$.

\item The $\mathtt{COLA}$ simulations for the case with the same stride $da=1/120$ but varying $z_{\rm init}$ as 29, 59, and 119, respectively, yield $\chi^2_{\rm min}$ values that are consistent with the $\chi^2$ distribution at the $2\sigma$ level ($\chi^2_{\rm min} < 27.3$ with DOF=16). Specifically, the $\chi^2_{\rm min}$ values are 20.8, 19.5 and 21.4 for the respective simulations, as listed in Tab.~\ref{tab:best_sigma_GADGET-COLA}.

\item In the $\mathtt{COLA}$ simulations for the case with $z_{\rm init}=29$ and varying $da$ values of $1/30$, $1/60$, and $1/120$, the resulting $\chi^2_{\rm min}$ values are consistently within the $2\sigma$ level. These values are 23.2, 18.0, and 20.8, respectively, as listed in Tab.~\ref{tab:best_sigma_GADGET-COLA}. This demonstrates the robustness of $\mathtt{COLA}$ across different $da$ values.
\end{itemize}

These findings suggest that in the case of using the $\mathtt{COLA}$ N-body method, when starting from the same initial redshift, the accuracy of the mock catalogue remains robust for different s. However, increasing the initial redshift does not automatically guarantee an improved accuracy of the mock catalogue. This phenomenon can be attributed to the accumulation of errors that occur during the numerous steps of low-force resolution computation, particularly in cases with a high initial redshift of simulation. On the other hand, choosing a very large $da$ (or a small $N_{\rm step}$) can also lead to inaccuracies in the simulation, since a very small number of time steps may fail to effectively capture the nonlinear evolution that occurs on small clustering scales. Therefore, selecting an appropriate combination of ($z_{\rm init}$,  $da$) is a critical factor in producing a high-quality mock catalog.

The main objective of this study is to provide an economical method for generating mock catalogues that can accurately reproduce the statistical properties of observed galaxies. Regarding Tab.~\ref{tab:best_sigma_GADGET-COLA}, the analyses indicate that, all tested cases are reasonable choices for the $\mathtt{COLA}$ simulation, as they can fit the observed data within a $2\sigma$ level. However, for the objectives of this study and considering the computational times listed in Tab.~\ref{tab:TestCosmo setting}, we select ($z_{\rm init}=29$, $da=1/30$) as our preferred $\mathtt{COLA}$ setting. This choice is based on the observation that there is no statistically significant difference among the cases with $da=1/30$, $1/60$, and $1/120$. By using fewer steps, it becomes easier to generate simulation catalogs efficiently and in large quantities, thereby in line with the ultimate goal of this work. In the following sections, through the analysis of various statistical metrics, we will demonstrate that this choice is indeed appropriate and yields results comparable to those obtained using $\mathtt{GADGET}$.

\begin{figure*}
    \centering
    \includegraphics[width=\textwidth]{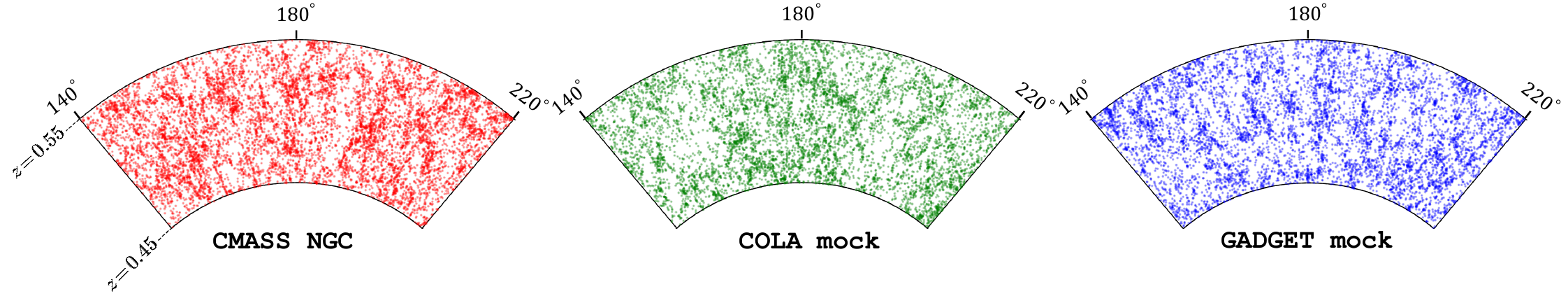}
    \caption{Pie plot of the $\mathtt{CMASS~NGC}$ data (left) and the light-cones of $\mathtt{COLA~mock}$ with ($z_{\rm init}=29,\;da=1/30$) (middle) and $\mathtt{GADGET~mock}$ (right), with a thickness of 4 deg (i.e., ${\rm DEC}\in[28^{\circ},32^{\circ}]$), with ${\rm RA}\in[140^{\circ}, 220^{\circ}]$). All shells represent $z\in[0.45, 0.55]$, corresponding to a comoving distance range of $r\in[1221.5, 1457.1],h^{-1}{\rm Mpc}$ in the fiducial cosmology.
    }
    \label{fig:sky-fan}
\end{figure*}
\begin{figure}
    \centering
    \includegraphics[width=0.45\columnwidth]{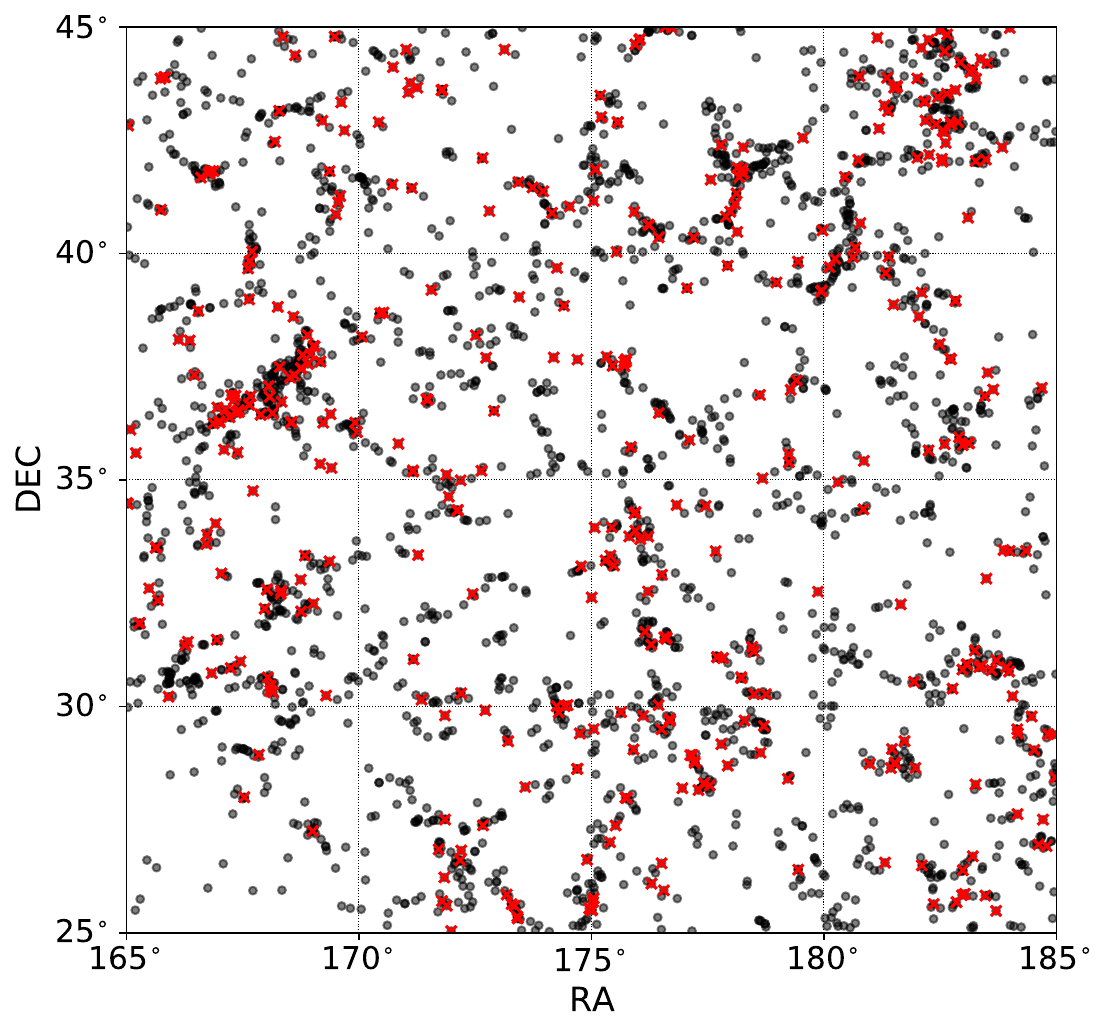}
    \caption{A redshift slice of $\mathtt{COLA~mock}$ ($z_{\rm init}=29,\;da=1/30$) with a mean redshift of $z_{\rm mean}=0.5$ and a bin width of $\Delta z=0.004$ in the sky region of $165^{\circ}<{\rm RA}<185^{\circ}$ and $25^{\circ}<{\rm DEC}<45^{\circ}$. Each side of this region extends approximately $468~h^{-1}{\rm Mpc}$ based on the fiducial cosmology. The gray points represent the full (sub)halos, while the red marks (``$\times$") correspond to the mock galaxies selected using the SHAM procedures.}
    \label{fig:TestSim_projmap}
\end{figure}

Using the best-fit $\sigma_{\rm scat}$ for different combinations ($z_{\rm init}$, $da$), one can employ our proposed procedure to generate mock catalogues. Visualization of $\mathtt{CMASS~NGC}$, $\mathtt{GADGET~mock}$, and $\mathtt{COLA~mock}$ ($z_{\rm init}=29$, $da=1/30$) is shown in Fig.~\ref{fig:sky-fan}, where we applied the same angular selection as $\mathtt{CMASS~NGC}$ to the mock catalogues. From this plot, it is evident that both the data and the mocks exhibit the same distribution, with no discernible visual differences apart from cosmic variance. Fig.~\ref{fig:TestSim_projmap} displays a redshift slice of the $\mathtt{COLA~mock}$ with ($z_{\rm init}=29$, $da=1/30$). The plot reveals the distribution of halos and catalogues selected using the SHAM process. Notably, it suggests that the $\mathtt{COLA~mock}$ catalogues demonstrate a more pronounced clustering feature, closely resembling that observed in galaxies. 

In summary, considering the incompleteness of the BOSS DR12 catalogue, by appropriately selecting the parameter $\sigma_{\rm scat}$ in the SHAM procedure described in Sect.~\ref{subsec:SHAM}, and choosing ($z_{\rm init}=29$, $da=1/30$) in $\mathtt{COLA}$, the statistical properties of $\mathtt{COLA~mocks}$ are similar to those of $\mathtt{GADGET~mock}$, and both of them are in good agreement with those of observed galaxies. Thus, our scheme greatly saves computational resources much more compared to the treatment using accurate N-body method, and this efficient mock generation method has broad applications in exploring the LSS of the Universe. For further verification, the subsequent section provides a thorough examination of the statistical properties for $\mathtt{COLA}$ simulations with varying $z_{\rm init}$ and $da$ values.

\section{Statistical measures of the mock catalogues}
\label{sec:Estimation of mocks}
In this section, we assess the fidelity of the mock catalogues by conducting a comprehensive comparison of various statistical measures derived from both the mock catalogues and the observational data. By examining these statistics, we can judge the quality and accuracy of the mock catalogues. Through this analysis, it will become evident that ($z_{\rm init}=29$, $da=1/30$) is the optimal choice for accurately reproducing the various statistical properties observed in the data, while minimizing computation time.

\subsection{The two-point clustering}
\label{subsec:Two-point clustering}

\begin{figure}
    \centering
    \includegraphics[width=0.7\columnwidth]{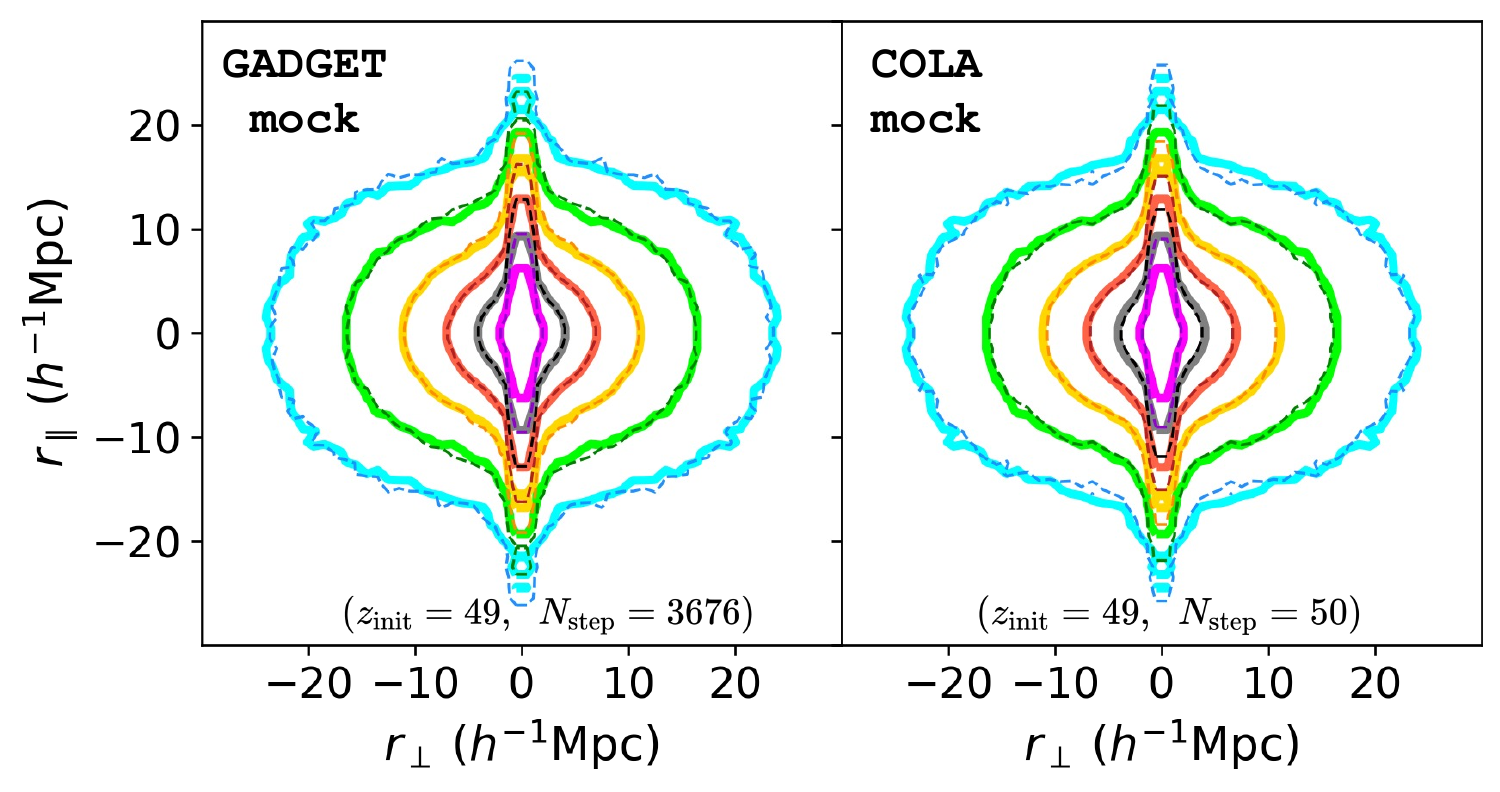}
    \caption{Contours of the correlation function $\xi(r_{\perp}, r_{\parallel})$ dependent on the separation parallel ($r_{\parallel}$) and perpendicular ($r_{\perp}$) to the LoS direction. The solid contours show the $\xi(r_{\perp}, r_{\parallel})$ of the $\mathtt{CMASS~NGC}$ catalogue in each panel. {\it Left}: comparison of $\xi(r_{\perp}, r_{\parallel})$ between the $\mathtt{GADGET~mock}$ (dashed) and $\mathtt{CMASS~NGC}$ (solid). {\it Right}: comparison of $\xi(r_{\perp}, r_{\parallel})$ between the $\mathtt{COLA~mock}$ with ($z_{\rm init}=49$, $da=1/50$, $N_{\rm step}=50$) (dashed) and $\mathtt{CMASS~NGC}$ (solid). The contours from the outside to the inside correspond to $\xi=(0.25, 0.5, 1, 2, 4, 8)$, respectively. It can be observed that the 2PCF of the $\mathtt{COLA}$ catalog is very similar to that predicted by $\mathtt{GADGET}$, and both of them are able to accurately reproduce the observational result.}
    \label{fig:TestSim_2Dxi}
\end{figure}

Based on $\xi(s,\mu)$, the 2PCF in $\xi(r_{\parallel}, r_{\perp})$ can be calculated by projecting the distances $s$ between the pair in LoS direction onto its parallel ($r_{\parallel}$) and perpendicular ($r_{\perp}$) components. In Fig.~\ref{fig:TestSim_2Dxi}, the left panel shows a comparison of the $\xi(r_{\parallel}, r_{\perp})$ contours between $\mathtt{CMASS~NGC}$ and $\mathtt{GADGET~mock}$ with ($z_{\rm init}=49$,  $N_{\rm step}=3676$). The right panel shows a comparison of the contours between $\mathtt{CMASS~NGC}$ and $\mathtt{COLA~mock}$ with ($z_{\rm init}=49$, $da=1/50$, $N_{\rm step}=50$). In each panel, we find that the Finger-of-God (FoG) effects~\citep{jackson1972critique} of the mock catalogues are slightly stronger than the observation data in the small $r_{\perp}$ range. The difference of the FoG effect between $\mathtt{GADGET~mock}$ and $\mathtt{COLA~mock}$ is small, indicating that the velocity distribution of $\mathtt{COLA~mock}$ approaches that of $\mathtt{GADGET~mock}$.

\begin{figure}
    \centering
    \includegraphics[width=\columnwidth]{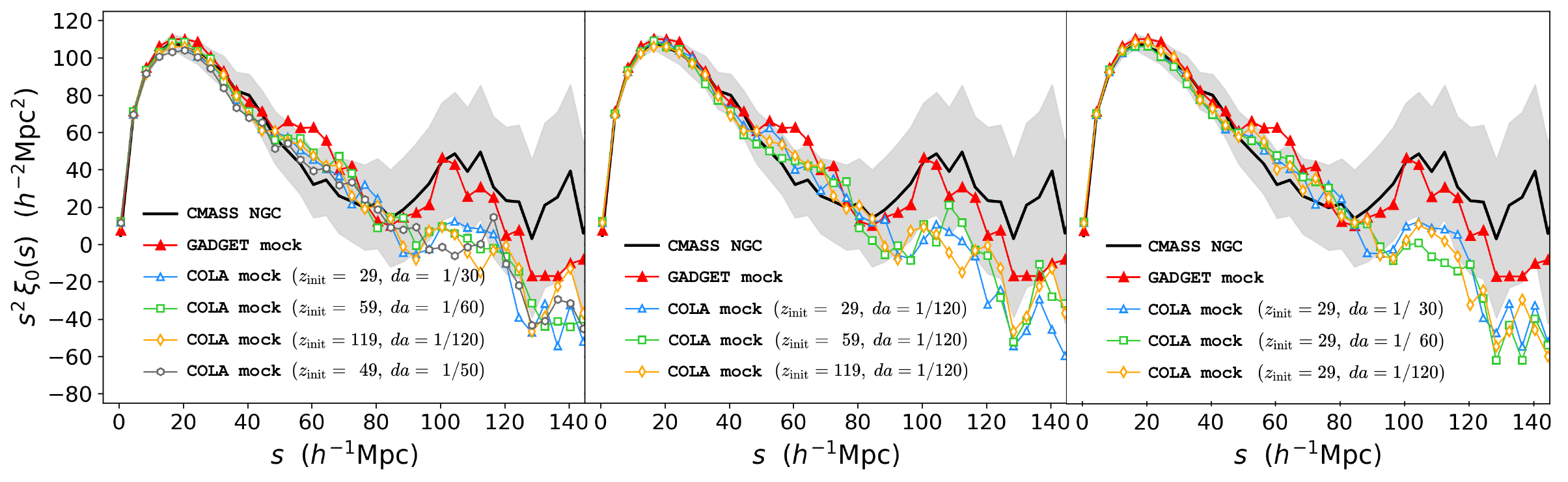}
    \caption{Monopole in configuration space for both the $\mathtt{CMASS~NGC}$ and the mock catalogues, using the optimal values of $\sigma_{\rm scat}$ obtained through the SHAM procedure. {\it Left}: the case of $\mathtt{COLA~mocks}$ with different $z_{\rm init}$, where $da$ is chosen as $da=1/(1+z_{\rm init})$. {\it Middle}: the case of $\mathtt{COLA~mocks}$ with different $z_{\rm init}$, while setting $da$ to a constant value of $1/120$. {\it Right}: the case of $\mathtt{COLA~mocks}$ with a fixed initial redshift of $z_{\rm init}=29$ while varying the value of $da$. In all panels, the black line corresponds to the monopole of the $\mathtt{CMASS~NGC}$ catalogue, with $2\sigma$ uncertainties (gray shaded) computed from 1000 $\mathtt{PATCHY~mocks}$. The monopoles of the $\mathtt{GADGET~mock}$ catalogue (solid triangle) and the $\mathtt{COLA~mocks}$ (hollow marked) are shown for comparison.}
    \label{fig:TestSim_xi1D}
\end{figure}

The resulting monopoles $\xi_0(s)$ for different mock catalogues are shown in Fig.~\ref{fig:TestSim_xi1D}. 
In the first case, with the setting of $da=1/(z_{\rm init}+1)$, the two-point statistical properties of $\mathtt{GADGET~mock}$ and $\mathtt{COLA~mock}$ for small $s$ values exhibit good consistency with the $\mathtt{CMASS~NGC}$ galaxies for various values of $z_{\rm init}$, including (29, 49, 59, 119). In the case of fixing $da=1/120$, the two-point statistical properties at small scales of both the $\mathtt{GADGET~mock}$ and $\mathtt{COLA~mock}$ with initial redshifts $z_{\rm init}= (29, 59, 119)$ also match well with those of $\mathtt{CMASS~NGC}$. In the last case, with the same initial redshift of $z_{\rm init}=29$, the two-point statistical properties at small scales of the $\mathtt{COLA~mock}$ with $da=(1/30,1/60,1/120)$ also agree with those of the $\mathtt{CMASS~NGC}$ galaxies. Moreover, in all three cases, noticeable deviations are observed at large scales due to the substantial sampling variance.

\subsection{The anisotropic two-point clustering}

To investigate the anisotropy using $\xi(s,\mu)$, we analyze the dependence of $\xi$ on $\mu$. Following the method presented in~\cite{li2015cosmological}, we integrate $\xi$ over the interval $s_{\rm min}\le s\le s_{\rm max}$, as given by 
\begin{align}
\label{eq:xi_Ds_mu}
\xi_{\Delta s}(\mu) \equiv \int_{s_{\rm min}}^{s_{\rm max}} \xi(s,\mu)ds \,.
\end{align}
Here a cut is applied on $\mu$ such that $\mu>\mu_{\rm min}$, to mitigate the geometric effect from the thin redshift shell of $0.45<z<0.55$. The quantity, $\xi_{\Delta s}(\mu)$, describes the angle dependence of the two-point clustering, allowing it to estimate the anisotropy of the galaxies. However, the value of $\xi$ at small scales is significantly affected by the FoG effect, which depends on the galaxy bias. This may result in a redshift evolution in $\xi_{\Delta s}(\mu)$, which is relatively difficult to model. On the other hand, at large scales, the measurement is dominated by noise due to poor statistics. According to~\cite{li2015cosmological}, the choice of $s_{\rm min} = 6\sim10~h^{-1} {\rm Mpc}$ and $s_{\rm max} = 40\sim70~h^{-1}{\rm Mpc}$ is appropriate, as it provides reliable, tight and unbiased constraints on cosmological parameters. For the aforementioned purposes, we have chosen $s_{\rm min} = 8~h^{-1}{\rm Mpc}$ and $s_{\rm max} = 60~h^{-1}{\rm Mpc}$, and and set $\mu_{\rm min}=0.12$ in this study. 

\begin{figure*}
    \centering
    \includegraphics[width=\columnwidth]{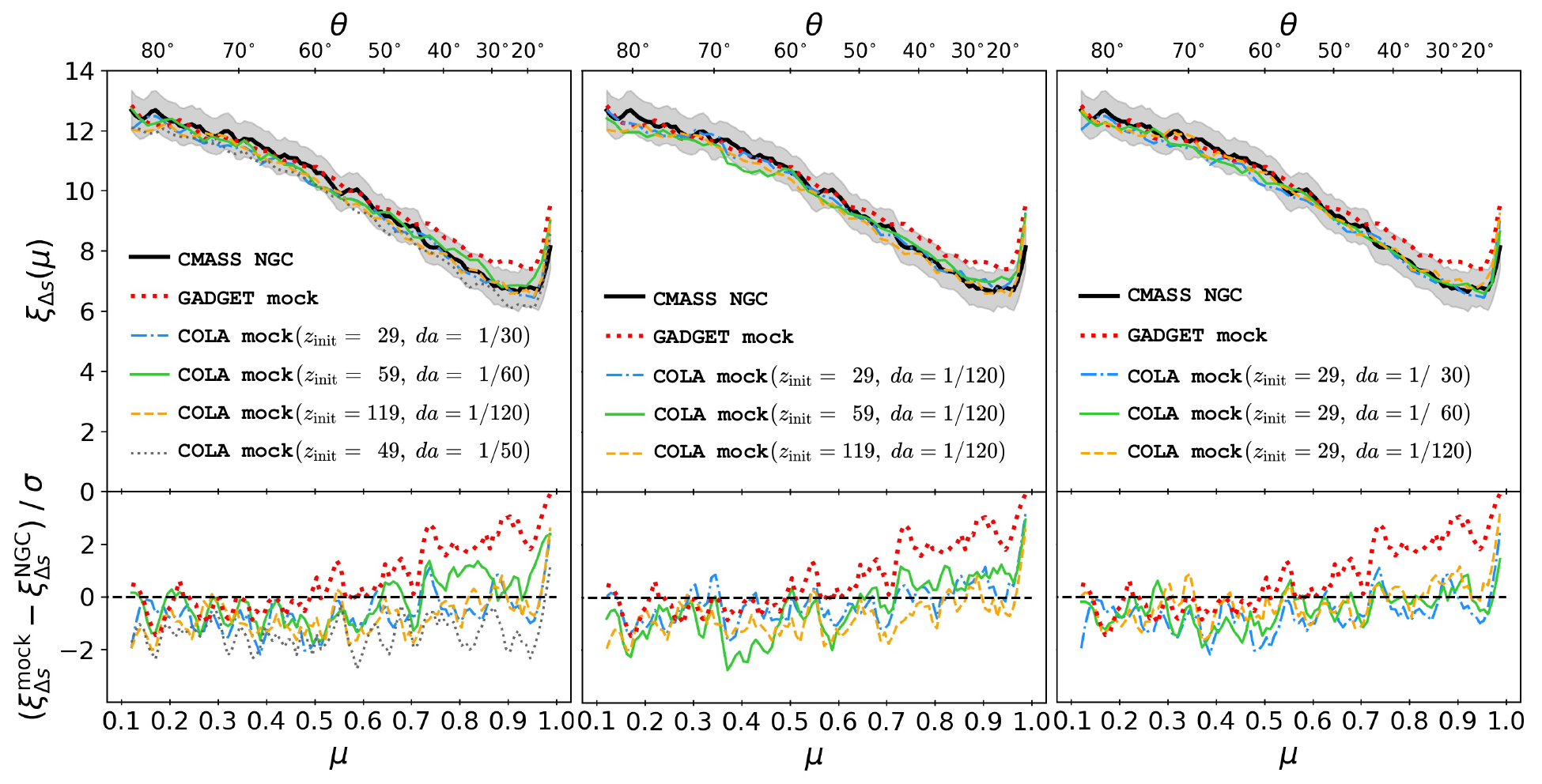}
    \caption{Angular dependence of $\xi$ on $\mu$ according to Eq.~\ref{eq:xi_Ds_mu}, with $s$ integrated out. The results are presented only for $\mu > \mu_{\rm min}$, where $\mu_{\rm min}=0.12$. For comparison, the gray shaded area represents the $2\sigma$ statistical uncertainty estimated using 1000 $\mathtt{PATCHY~mocks}$. The bottom panels are the relative errors between the mock catalogs and the observation with respect to the $1\sigma$ statistical uncertainty, defined by $(\xi_{\Delta s}^{\rm mock}-\xi_{\Delta s}^{\rm NGC})/\sigma$. The angular dependence for $\mathtt{GADGET~mock}$ and the three different $\mathtt{COLA}$ settings are displayed from left to right for comparison.}
\label{fig:TestSim_ximu}
\end{figure*}

In Fig.~\ref{fig:TestSim_ximu}, the anisotropic clustering patterns of the $\mathtt{CMASS~NGC}$, $\mathtt{GADGET~mock}$, and $\mathtt{COLA~mocks}$ are compared using the $\xi_{\Delta s}(\mu)$ statistic. The $\mathtt{GADGET~mock}$ catalogue has a similar anisotropy pattern to the observational result, with a relative error that falls within the $2\sigma$ uncertainties. Fig.~\ref{fig:TestSim_ximu} also shows the angular dependence of $\xi$ on $\mu$ for the three different $\mathtt{COLA}$ settings. In the first case (left), we vary $z_{\rm init}$ while adjusting $da$ according to $da=1/(1+z_{\rm init})$. In the second case (middle), we increase $z_{\rm init}$ from 29 to 119 while keeping a fixed $da$ of $1/120$. We find consistent clustering patterns with that of $\mathtt{CMASS~NGC}$ at $2\sigma$ level. Similarly, in the third case (right), where $z_{\rm init}$ is fixed at 29 and $da$ is varied, the $\mathtt{COLA~mock}$ catalogues all agree with the observed data in terms of the statistical uncertainty. This observation suggest that the $\mathtt{COLA}$ simulation maintains good robustness with respect to changes in $da$.

\subsection{Three-point clustering}
Furthermore, we can use the three-point clustering analysis to assess the accuracy of our mock data. The three-point correlation function (3PCF) can be expressed as the probability of finding a galaxy in each of the volume elements $dV_1$, $dV_2$, and $dV_3$, given that these elements are arranged in a particular configuration defined by the sides of a triangle ($\bm{r}_1$, $\bm{r}_2$, and $\bm{r}_3$). This joint probability can be mathematically represented as:
\begin{equation}
\begin{aligned}
d P_{1,2,3}=\bar{n}^3[1 & +\xi\left(\bm{r}_1\right)+\xi\left(\bm{r}_2\right)+\xi\left(\bm{r}_3\right)+ \\
& \left.+\zeta\left(\bm{r}_1, \bm{r}_2, \bm{r}_3\right)\right] d V_1 d V_2 d V_3\,.
\end{aligned}
\end{equation}
The above expression consists of several components: the sum of 2PCFs for each side of the triangle, the complete 3PCF denoted by $\xi$, and $\bar{n}$ which represents the mean density of data points. The 3PCF estimator is given by~\citep{szapudi1998new},
\begin{align}
    \label{eq:3pCF_definition}
    \zeta=\frac{DDD-3DDR+3DRR-RRR}{RRR}\,,
\end{align}
where each term represents the normalized count of triplets in the data ($D$) and random ($R$) samples that satisfy a specific triangular configuration of our choice. The function $\zeta$ in the isotropic case depends on the three sides of the triangle, denoted as $(r_1, r_2, r_3)$. In contrast, in the anisotropic case, $\zeta$ depends on the three sides of the triangle as well as two angles ($\theta_1$ and $\theta_2$) relative to the LoS direction ($\hat{\bm{z}}$), where $\cos \theta_1= \hat{\bm{r}}_1\cdot \hat{\bm{z}}$ and $\cos\theta_2=\hat{\bm{r}}_2\cdot \hat{\bm{z}}$. Although one could perform an analysis in the anisotropic 5-parameter space $(r_1, r_2, r_3, \theta_1, \theta_2)$, this study only considers the isotropic case of 3PCF, reducing 5-parameter space  to $\zeta(r_1, r_2, r_3)$. We utilize the GRAMSCI\footnote{https://bitbucket.org/csabiu/gramsci} (GRAph Made Statistics for Cosmological Information) method, a publicly available code developed by~\cite{sabiu2019graph}, to estimate the three-point statistics $\zeta(r_1, r_2, r_3)$. This method combines the concepts of KD-trees and graph databases to speed up the calculation of 3PCF, making it more efficient than traditional methods.

\begin{figure*}
    \centering
    \includegraphics[width=\columnwidth]{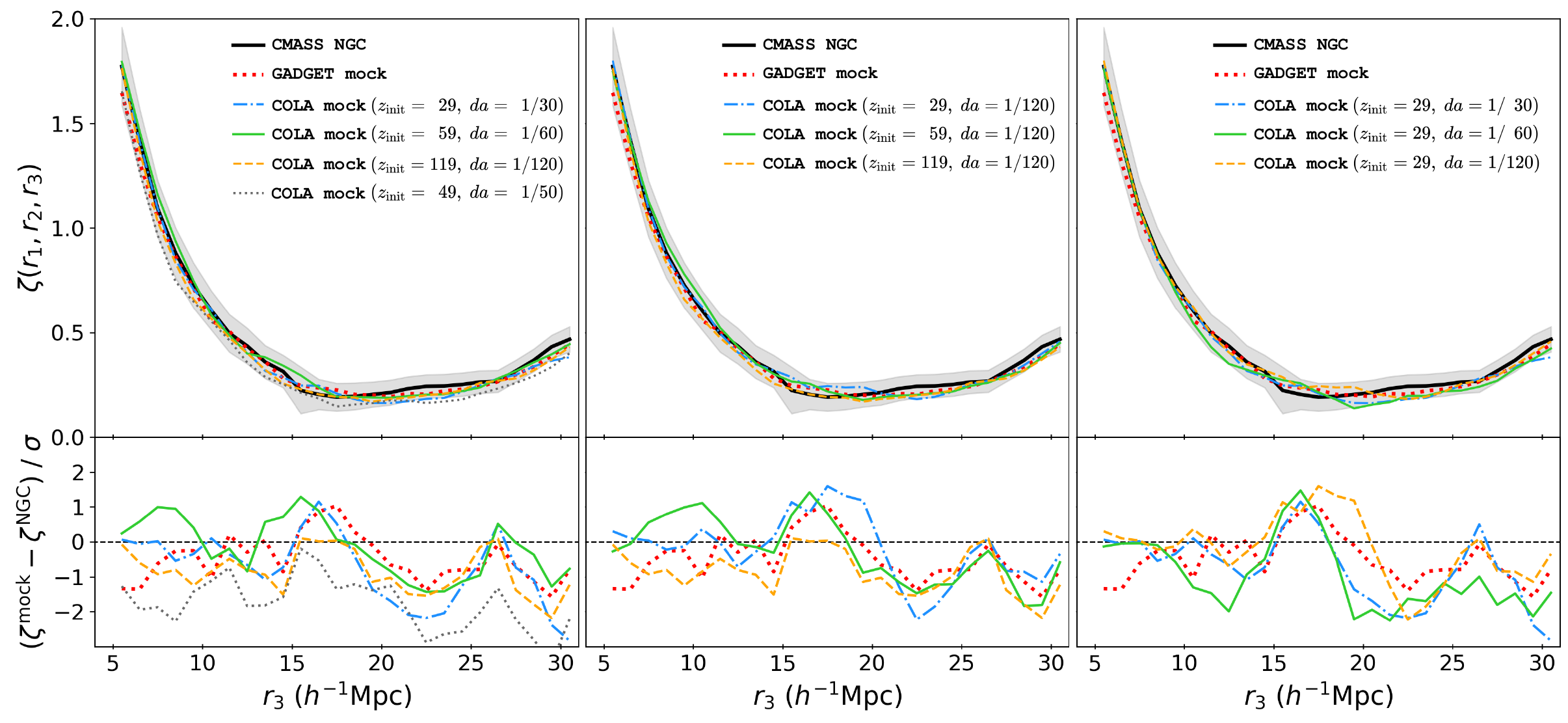}
    \caption{Comparison of three-point statistics $\zeta(r_{1}, r_{2}, r_{3})$ between the observation data $\mathtt{CMASS~NGC}$ and mock catalogs. The shaded area in the top panels represents the $2\sigma$ uncertainties originating from the initial randomness, which are estimated using $\mathtt{PATCHY~mock}$. $r_{1}$ and $r_{2}$ are both fixed at 15.5 $h^{-1}{\rm Mpc}$, and $r_{3}$ varies from 5 to 31 $h^{-1}{\rm Mpc}$.The bottom panels show the relative errors between the mock catalogs and $\mathtt{CMASS~NGC}$ with respect to the 1$\sigma$ statistical uncertainty, defined by $(\zeta^{\rm mock}-\zeta^{\rm NGC})/\sigma$. Same as in Fig.~\ref{fig:TestSim_ximu}, three different $\mathtt{COLA}$ settings are used for comparison.}
    \label{fig:TestSim_3pcf}
\end{figure*}

The resulting 3PCF, as shown in Fig.~\ref{fig:TestSim_3pcf}, indicates that the mock catalogues have comparable three-point clustering with that observed in $\mathtt{CMASS~NGC}$. In the top panels, the solid line represents the resulting 3PCF from the observed $\mathtt{CMASS~NGC}$ galaxies, and the shaded area represents the $2\sigma$ uncertainties obtained from  $\mathtt{PATCHY~mocks}$. Based on the three-point clustering, the $\mathtt{GADGET~mock}$ shows a good match with the observed data. For comparison, we illustrate the three different $\mathtt{COLA}$ settings from left to right. The adopted values of $z_{\rm init}$ and $da$ are listed in each panel. In most of the adopted parameter combinations for ($z_{\rm init}$, $da$), we observe that the resulting 3PCFs are consistent with those of the observed data from $\mathtt{CMASS~NGC}$ within the $2\sigma$ level. Therefore for saving computing resources, ($z_{\rm init}=29$, $da=1/30$) is the optimal option for achieving a good consistency with the observed galaxies in terms of 3PCF.

\subsection{Power spectrum multipoles}
\label{sect.pk}
In order to estimate the redshift-space clustering of mock catalogues, we apply the FFT-based anisotropic power spectrum estimator proposed by \cite{hand2017optimal}. Starting with the weighted object density field $F(\mathbf{r})$~\citep{feldman1993power}, defined as
\begin{align}
    \label{eq:F(r) the weighted object density field}
    F(\mathbf{r}) = \frac{w(\mathbf{r})}{I^{1/2}}\left[ n(\mathbf{r})-\alpha n_{s}(\mathbf{r})\right], 
\end{align}
where $n$ and $n_{s}$ are the number density field for the catalogue of the observed objects and synthetic catalogue of random objects, $w(\mathbf{r})$ is a general weighting scheme and the factor $\alpha$ normalizes the synthetic catalogue to the number density of the observed objects. $I$ is the factor for normalization, defined as $I=\int d\mathbf{r}~ \Bar{n}^{2}(\mathbf{r}) w^{2}(\mathbf{r})$. Without loss of generality, we choose the FKP weights~\citep{feldman1993power} for both the data and random. The FKP method is an optimal approach that accommodates variations in density across a survey, effectively balancing sample variance and shot noise. Thus, each galaxy is assigned a weight of $w=w_{\rm FKP}(\mathbf{r}) \equiv\left(1+n(\mathbf{r}) P_0\right)^{-1}$, and we adopt $P_0=10^{4}~h^{-3}{\rm Mpc}^{3}$ as the fiducial value for a typical galaxy survey.

The FFT-based anisotropic power spectrum estimator can be defined as
\begin{align}
    P_{\ell}(k) = \frac{2\ell+1}{I} \int \frac{d\Omega_{k}}{4\pi}\left[F_{0}(\bm{k})F_{\ell}(-\bm{k}) -P^{\rm noise}_{\ell}(\mathbf{k})\right]\,,
\end{align}
where $\Omega_k$ is the solid angle of the Fourier mode. $P_{\ell}^{\rm noise}$ is the shot noise, given by
\begin{align}
    \label{eq:the_shot_noise}
    P_{\ell}^{\rm noise}(\mathbf{k}) = (1+\alpha) \int d\mathbf{r}\; \Bar{n}(\mathbf{r})\;w^{2}(\mathbf{r}) \;\mathcal{L}_{\ell}(\hat{\mathbf{k}}\cdot\hat{\mathbf{r}}) 
\end{align}
where $\mathcal{L}_{\ell}$ is the $\ell$-th order Legendre polynomial, and one can assume that $P_{\ell}^{\rm noise}=0$ for $\ell>0$. $F_{\ell}(\bm{k})$ is the Legendre expansion of the weighted object density field $F(\mathbf{r})$ in Fourier space, given by
\begin{align}
F_{\ell}(\mathbf{k})=\int d^3 r F(\mathbf{r}) \mathcal{L}_{\ell}(\hat{\mathbf{k}} \cdot \hat{\mathbf{r}}) e^{i \mathbf{k} \cdot \mathbf{r}}.
\end{align}
 
Utilizing the NBODYKIT\footnote{https://github.com/bccp/nbodykit} library~\citep{hand2018nbodykit}, we calculate the power spectrum multipoles $P_{\ell}(k)$. In this study, we consider only the multipoles of $\ell = 0, 2$, and $4$, as higher orders are mostly dominated by noise modes. As mentioned earlier in Fig.~\ref{fig:sky-region}, we do apply the same angular selection of $\mathtt{CMASS~NGC}$ to the mock catalogues when calculating the multipoles $P_{\ell}(k)$. For all power spectrum measurements, we perform a linear binning in the chosen Fourier space with a bin width of $\Delta k= 0.01~h/{\rm Mpc}$, and the range of the Fourier space chosen for binning is $k\in[0.002,0.4]~h/{\rm Mpc}$.

\begin{figure*}
\centering
    \includegraphics[width=\textwidth]{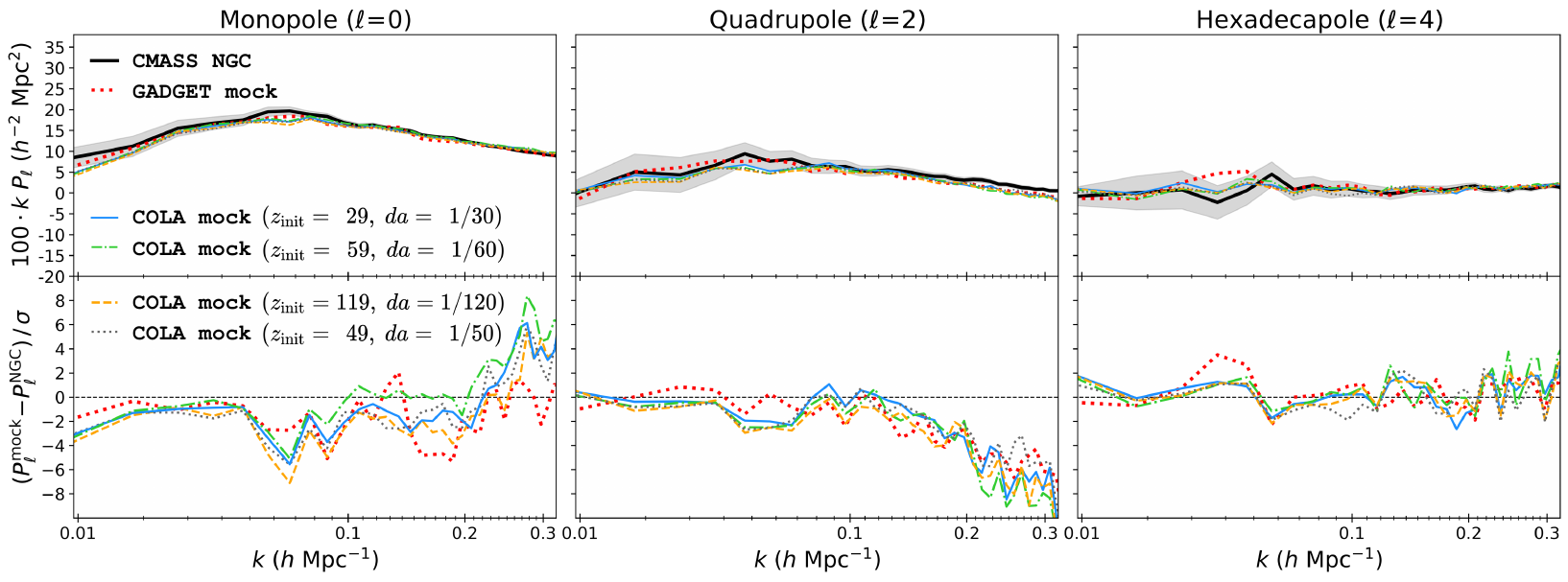}
    \caption{{\it Top}: comparison of power spectrum multipoles of $\ell=0,~2,~4$ for $\mathtt{CMASS~NGC}$ and $\mathtt{GADGET~mock}$ as well as $\mathtt{COLA~mock}$ with $z_{\rm init}=29,~49,~59,~119$ when the time stride $da=1/N_{\rm step}$. The shaded region represents the $2\sigma$ statistical uncertainty obtained from measuring 1000 $\mathtt{PATCHY~mocks}$ for each multipole. {\it Bottom}: relative error between the mocks and $\mathtt{CMASS~NGC}$ with respect to the $1\sigma$ statistical uncertainty, defined by $(P_{\ell}^{\rm mock}-P_{\ell}^{\rm NGC})/\sigma$ for $\ell=0,~2,~4$.}
\label{fig:TestSim_diff-Zinit_da1-Nstep_multipoles}
\end{figure*}

\begin{figure*}
\centering
    \includegraphics[width=\textwidth]{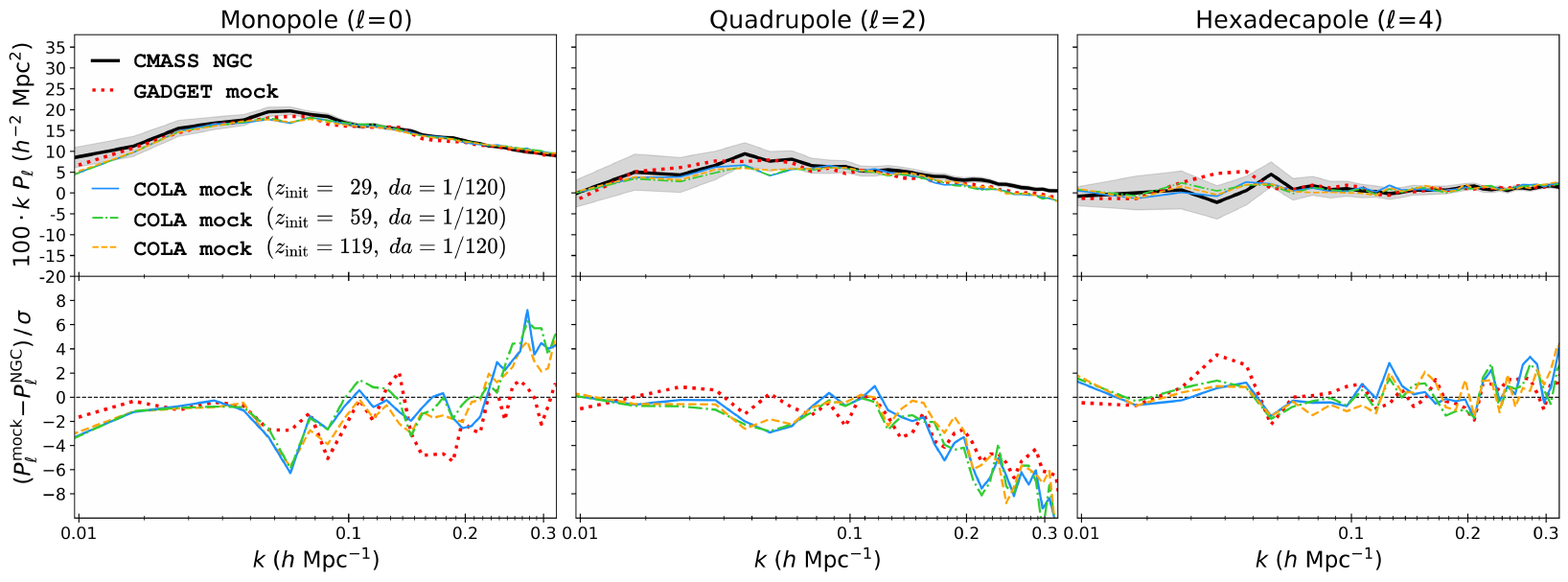}
    \caption{{\it Top}: comparison of power spectrum multipoles of $\ell=0,~2,~4$ for $\mathtt{CMASS~NGC}$ and $\mathtt{GADGET~mock}$ as well as $\mathtt{COLA~mock}$ with $z_{\rm init}=29,~59,~119$ when $da=1/120$. The shaded region represents the $2\sigma$ statistical uncertainty obtained from measuring 1000 $\mathtt{PATCHY~mocks}$ for each multipole. {\it Bottom}: relative error between the mocks and $\mathtt{CMASS~NGC}$ with respect to the $1\sigma$ statistical uncertainty, defined by $(P_{\ell}^{\rm mock}-P_{\ell}^{\rm NGC})/\sigma$ for $\ell=0,~2,~4$.}
\label{fig:TestSim_diff-Zinit_da1-120_multipoles}
\end{figure*}

\begin{figure*}
\centering
    \includegraphics[width=\textwidth]{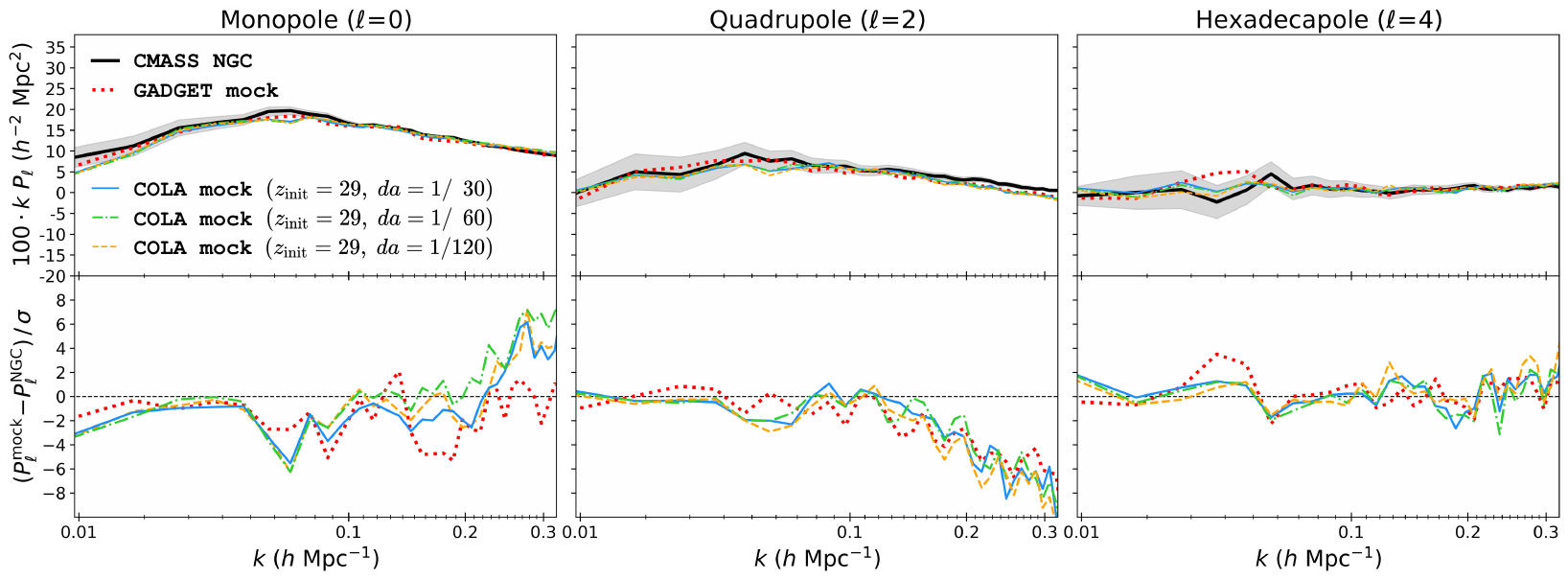}
    \caption{{\it Top}: comparison of power spectrum multipoles of $\ell=0,~2,~4$ for $\mathtt{CMASS~NGC}$ and $\mathtt{GADGET~mock}$ as well as $\mathtt{COLA~mock}$ with $da=1/30,~1/60,~1/120$ when $z_{\rm init}=29$. The shaded region represents the $2\sigma$ statistical uncertainty obtained from measuring 1000 $\mathtt{PATCHY~mocks}$ for each multipole. {\it Bottom}: relative error between the mocks and $\mathtt{CMASS~NGC}$ with respect to the $1\sigma$ statistical uncertainty, defined by $(P_{\ell}^{\rm mock}-P_{\ell}^{\rm NGC})/\sigma$ for $\ell=0,~2,~4$.}
\label{fig:TestSim_Zinit-29_da1-Nstep_multipoles}
\end{figure*}

Fig.~\ref{fig:TestSim_diff-Zinit_da1-Nstep_multipoles}, \ref{fig:TestSim_diff-Zinit_da1-120_multipoles} and \ref{fig:TestSim_Zinit-29_da1-Nstep_multipoles} display the comparisons of power spectrum multipoles ($\ell=0,~2,~4$) for $\mathtt{CMASS~NGC}$, $\mathtt{GADGET~mock}$, and $\mathtt{COLA~mock}$ with three cases of ($z_{\rm init}$, $da$) respectively. As shown in the three figures, the black solid lines represent the results from $\mathtt{CMASS~NGC}$, while the shaded region represents the 2$\sigma$ statistical uncertainties obtained from 1000 $\mathtt{PATCHY~mocks}$. 

As observed, the monopole $P_0(k)$, which represents the angular averaged power spectrum, is displayed in the left panels of Fig.~\ref{fig:TestSim_diff-Zinit_da1-Nstep_multipoles}, \ref{fig:TestSim_diff-Zinit_da1-120_multipoles} and \ref{fig:TestSim_Zinit-29_da1-Nstep_multipoles}. At large scales, when $k<0.1h/{\rm Mpc}$, most of the points for $P_0(k)$ of $\mathtt{GADGET~mock}$ and $\mathtt{COLA~mock}$ have a good consistency with that of $\mathtt{CMASS~NGC}$ within $2\sigma$ level. The $\mathtt{GADGET~mock}$ and $\mathtt{COLA~mock}$ demonstrate a comparable ability to reproduce $P_0(k)$ to that of $\mathtt{CMASS~NGC}$. Moreover, when $k\in[0.1, 0.2]~h/{\rm Mpc}$, the monopole of $\mathtt{COLA~mock}$ approaches the $\mathtt{CMASS~NGC}$ measurement even more closely than that of $\mathtt{GADGET~mock}$. In addition, at small scales, $k>0.2~h/{\rm Mpc}$, $P_{0}(k)$ of $\mathtt{COLA~mock}$ gradually deviates from the real data measurement and the theoretical prediction beyond the $2\sigma$ level. This deviation may be due to the relatively low force resolution of the $\mathtt{COLA}$ N-body method.

In addition to the monopole, the middle panels of Fig.~\ref{fig:TestSim_diff-Zinit_da1-Nstep_multipoles}, \ref{fig:TestSim_diff-Zinit_da1-120_multipoles} and \ref{fig:TestSim_Zinit-29_da1-Nstep_multipoles} present the results for the quadrupole, $P_{2}(k)$, which represents the leading anisotropies in the power spectrum in redshift space. At large scales, all of mocks are well consistent with observations within a $2\sigma$ uncertainty. However, for $k>0.2~h/{\rm Mpc}$, both $\mathtt{GADGET~mock}$ and $\mathtt{COLA~mock}$ have lower $P_{2}(k)$ values compared with $\mathtt{CMASS~NGC}$. This is likely due to the fact that both simulation mocks cannot reproduce the real RSD effects well at small scales, which is caused by the relatively low resolution of the simulations with only $1024^{3}$ particles in a box with a side length of $800~h^{-1}{\rm Mpc}$.

Besides, the right panels of  Fig.~\ref{fig:TestSim_diff-Zinit_da1-Nstep_multipoles}, \ref{fig:TestSim_diff-Zinit_da1-120_multipoles} and \ref{fig:TestSim_Zinit-29_da1-Nstep_multipoles} show the next-to-leading anisotropies represented by the hexadecapole $P_4(k)$. It can be observed that when $k<0.3~h/{\rm Mpc}$, the $P_{4}(k)$ values from $\mathtt{CMASS~NGC}$, $\mathtt{GADGET~mock}$, and $\mathtt{COLA~mock}$ are very similar, with most of the values falling within the $2\sigma$ region, indicating good consistency with the theoretical prediction. 

In summary, based on the analysis of multipoles presented above, it can be concluded that $\mathtt{COLA~mock}$ catalogs with the used combinations of ($z_{\rm init}$, $da$) perform similarly to $\mathtt{GADGET~mock}$ and can accurately reproduce the power spectra of anisotropies in most cases.

\subsection{Performance test of varying cosmological parameters}
\label{subsec:Mock set}

To further assess the impact of different cosmological parameters on our proposed scheme and to validate its performance over various cosmological models, we varied the values of cosmological parameters in the $\mathtt{COLA}$ simulations. Using the same SHAM procedure as described in Sect.~\ref{subsec:SHAM}, we generated $\mathtt{COLA~mocks}$ for each set of cosmological parameters, with the setting of ($z_{\rm init}=29$, $da=1/30$). Based on these mock catalogs and their corresponding 2PCFs, we determined the optimal value of the parameter $\sigma_{\rm scat}$ by minimizing the $\chi^2$ of the 2PCFs through Eq.~\ref{eq:chi2_for_HAM_best_scatter}, and the corresponding results are shown in Tab.~\ref{tab:MultiCosmo setting}. 

In this section, we select the base cosmology with the following parameter values: $\Omega_{m}=0.31$, $\Omega_{b}=0.048$, $\Omega_\Lambda=0.69$, $w=-1.0$, $\sigma_{8}=0.82$, $h=0.68$. These values closely approximate the mean constraint derived from the Planck 2015 results~\citep{ade2016planck}. It is worth noting that this chosen cosmology differs from the one used in the previous sections of this study. In particular, we varied the values of $\Omega_{m}$, $w$, and $\sigma_8$ with respect to the base cosmological model, resulting in slight changes within the ranges of $\Omega_{m}\in[0.29,0.30]$, $w\in[-1.1,-0.8]$, and $\sigma_8\in[0.81,0.84]$. 
As indicated in Tab.~\ref{tab:MultiCosmo setting}, the best-fit value of $\sigma_{\rm scat}$ shifts from 0.50 for the base cosmological model to a range of $[0.47, 0.62]$ when different cosmological parameters are varied. This indicates that the values of $\sigma_{\rm scat}$ do not change significantly, suggesting that the estimate is robust against different cosmological models.

The relative changes of $\chi^2_{\rm min}$ with respect to the statistical uncertainty, defined as $\Delta \chi^2_{\rm min}/\sigma$ where $\Delta \chi^2_{\rm min}\equiv \chi^2_{\rm min} - \chi^2_{\rm min}({\rm base})$, are listed in Tab.~\ref{tab:MultiCosmo setting}. Note that, the cosmological parameters chosen in the case ``Base" are $2\sigma$ consistent with the Planck 2015 cosmology results~\citep{Planck:2015fie}, whereas they are slightly different from what we used in the previous sections (see the detailed values in Sect.~\ref{subsec:mock production}).   

Since ${\rm DOF}=16$ for the fit, according to the $\chi^2$ distribution, the expected standard deviation ($\sigma$) is approximately 5.66, calculated as $\sigma \approx \sqrt{2 \times {\rm DOF}}$. As shown in Tab.~\ref{tab:MultiCosmo setting}, for all cases, the absolute values of the changes in $\chi_{\rm min}^2$ are smaller than the 2$\sigma$ uncertainty. These results further support the validity of our proposed scheme for generating mock data for various cosmological models.

\begin{table}
    \centering
    \renewcommand\arraystretch{1.4}
    \begin{tabular}{c|ccc||cc}
        \hline\hline
        Name & $\Omega_{m}$ & $w$ & $\sigma_{8}$ & $\sigma_{\rm scat}^{\rm best}$ & $\Delta \chi_{\rm min}^2/\sigma$ \\ \hline
        Base & 0.31 & -1.0 & 0.82 & 0.50 & 0.00 \\ \hline 
        $\Omega_{m}^{-}$  & \textbf{0.30} & -1.0 & 0.82 & 0.59 &  0.76 \\ \hline 
        $\Omega_{m}^{+}$  & \textbf{0.32} & -1.0 & 0.82 & 0.55 &  1.23 \\ \hline  
        $\Omega_{m}^{--}$ & \textbf{0.29} & -1.0 & 0.82 & 0.62 &  0.53 \\ \hline  
        $\Omega_{m}^{++}$ & \textbf{0.33} & -1.0 & 0.82 & 0.47 & -1.31 \\ \hline  
        
        $w^{-}$               & 0.31 & \textbf{-1.1} & 0.82 & 0.49 & -1.13 \\ \hline  
        $w^{+}$               & 0.31 & \textbf{-0.9} & 0.82 & 0.50 &  0.42 \\ \hline  
        $w^{--}$              & 0.31 & \textbf{-1.2} & 0.82 & 0.49 &  0.49 \\ \hline  
        $w^{++}$              & 0.31 & \textbf{-0.8} & 0.82 & 0.50 &  0.03 \\ \hline  
        
        $\sigma_{8}^{-}$      & 0.31 & -1.0 & \textbf{0.81} & 0.55 &  0.49 \\ \hline  
        $\sigma_{8}^{+}$      & 0.31 & -1.0 & \textbf{0.83} & 0.54 &  0.20 \\ \hline  
        $\sigma_{8}^{--}$     & 0.31 & -1.0 & \textbf{0.80} & 0.48 &  1.00 \\ \hline  
        $\sigma_{8}^{++}$     & 0.31 & -1.0 & \textbf{0.84} & 0.49 & -0.81 \\ \hline\hline
        
    \end{tabular}
    \caption{Performance test of $\mathtt{COLA~mocks}$ with the setting of ($z_{\rm init}=29$, $da=1/30$) by evaluating the effects of varying different cosmological parameters. When a cosmological parameter is changed from the base cosmology, it is written in bold. We primarily focus on varying three parameters: $\Omega_m$ (specific values listed in the 2nd column), $w$ (the 3rd column), and $\sigma_8$ (the 4th column). Depending on the cosmological parameters used in $\mathtt{COLA}$ simulation, the distribution of the mock catalogue generated by our SHAM process exhibits slight variations. By minimizing the $\chi^2$ of the 2PCF (based on Eq.~\ref{eq:chi2_for_HAM_best_scatter}), we determined the optimal value of $\sigma_{\rm scat}^{\rm best}$ (the 5th column) that provides the best fit between the simulated and observed 2PCF.  In all cases, the optimal values of $\sigma_{\rm scat}^{\rm best}$  do not change significantly compared with that in the base cosmology despite significant variations in the cosmological parameters. The relative changes of $\chi^2_{\rm min}$ with respect to the statistical uncertainty are presented in the rightmost column, where $\Delta\chi^2_{\rm min} \equiv \chi^2_{\rm min} - \chi^2_{\rm min}({\rm base})$. This implies that the relative changes for the different cosmology models are not significantly different from the $2\sigma$ uncertainty, where $\sigma\approx\sqrt{2\times \rm DOF}\simeq5.66$ with ${\rm DOF}=16$. These catalogues can be accessed through at this url:~\href{https://nadc.china-vo.org/res/r101298/}{
\nolinkurl{https://nadc.china-vo.org/res/r101298/}
    }. }
    \label{tab:MultiCosmo setting}
\end{table}

\section{Conclusions}
\label{sec:conclusions}

Fast mock generation has wide-ranging applications in studying LSS of the Universe and effectively evaluating various theoretical models on cosmological scales in Stage IV surveys. This is particularly crucial for approaches that necessitate numerous mock catalogues, such as machine learning. In this study, we utilized $\mathtt{COLA}$ simulations to propose an efficient method for generating mock catalogues rapidly. We have confirmed that the mocks generated through $\mathtt{COLA}$ accurately reproduce the observed statistical properties of LSS using various statistical estimators, including 2PCF and 3PCF in redshift space, and the power spectrum multipoles. Meanwhile, we find that the performance of $\mathtt{COLA~mocks}$ is comparable to that of $\mathtt{GADGET}$ simulations.

There are only three free parameters involved in generating $\mathtt{COLA~mocks}$: $z_{\rm init}$, $da$ and $\sigma_{\rm scat}$. $\sigma_{\rm scat}$ is used for the SHAM procedure, while $z_{\rm init}$ represents the initial redshift of $\mathtt{COLA}$ simulation and $da$ is the time stride in $\mathtt{COLA}$. Smaller values of $da$ lead to more time-consuming simulations. Using the two-point clustering statistic, we computed the $\chi^2$ values between the mocks and the observed galaxies (BOSS DR12 $\mathtt{CMASS~NGC}$ catalogue). By minimizing $\chi^2$, we obtained the optimal value of $\sigma_{\rm scat}$ for a given combination of ($z_{\rm init}$, $da$). 

Based on the tests conducted by varying the combination of ($z_{\rm init}$, $da$), we have found that, in most cases, $\mathtt{COLA}$ can yield good consistency with various statistical quantities of observed galaxies. Especially, in the case of setting $z_{\rm init}$ to 29, 59, and 119, respectively, while keeping $da$ constant (with a chosen value of $da = 1/120$), based on the isotropic results, the anisotropic two-point clustering, and the three-point clustering, as well as the power spectrum multipoles, we observe that the statistical properties of these resulting $\mathtt{COLA~mocks}$ are consistent with each other. Therefore, the simulated $\mathtt{COLA}$ method demonstrates robustness when varying the initial redshift, while the time stride $da$ remains fixed. Furthermore, taking into account the computational cost of each simulation setup, the optimal choice is ($z_{\rm init}=29$, $da=1/30$). This choice not only allows for fast mock generation that statistically matches observations, but also minimizes computational demands.

Furthermore, we evaluated the performance of $\mathtt{COLA~mocks}$ over different cosmological models. By considerably varying $\Omega_m$, $w$, and $\sigma_8$, we observe that the best-fit values of $\sigma_{\rm scat}$ do not change significantly. Additionally, the changes in the $\chi^2$ values for 2PCF are well below the $2\sigma$ uncertainty. These results indicate the validity and robustness of our fast mock generation scheme across different cosmological models. 

In summary, our scheme successfully produces mock catalogs that possess similar statistical properties to the observed galaxies. In the future, we plan to generate lightcone catalogs spanning a broad range of redshifts, fulfilling the data requirements for most cosmological studies, instead of focusing on a narrow redshift bin as done in this study.


\section*{Acknowledgments}
We thank Cristiano G. Sabiu for helpful discussions and the support of GRAMSCI. This work is supported by National SKA Program of China (2020SKA0110401, 2020SKA0110402, 2020SKA0110100), the National Key R\&D Program of China (2020YFC2201600), National Science Foundation of China (11890691, 12203107), the China Manned Space Project with No. CMS-CSST-2021 (A02, A03, B01), the Science and Technology Program of Guangzhou (202002030360), the Guangdong Basic and Applied Basic Research Foundation (2019A1515111098), and the 111 project of the Ministry of Education No. B20019. We also wish to acknowledge the Beijing Super Cloud Center (BSCC) and Beijing Beilong Super Cloud Computing Co., Ltd (\url{http://www.blsc.cn/}) for providing HPC resources that have significantly contributed to the research results presented in this paper.

%

\vspace{5mm}


\software{CAMB~\citep{lewis2000efficient},
          COLA~\citep{Tassev_2013},  
          ROCKSTAR~\citep{behroozi2012rockstar}, 
          Consistent-Trees~\citep{behroozi2012gravitationally},
          NBODYKIT~\citep{hand2018nbodykit},
          CUTE~\citep{2015ascl.soft05016A},
          GRAMSCI~\citep{sabiu2019graph},
          NumPy~\citep{harris2020array},
          SciPy~\citep{virtanen2020scipy},
          Astropy~\citep{robitaille2013astropy},
          h5py~\citep{collette_python_hdf5_2014},
          mpi4py~\citep{dalcin2021mpi4py},
          pandas~\citep{the_pandas_development_team_2023_10045529,mckinney2010data},
          Matplotlib~\citep{hunter2007matplotlib}, 
          OpenMPI~\citep{gabriel2004open}
          }


\bibliography{sample631}{}
\bibliographystyle{aasjournal}

\end{document}